\documentclass[lettersize,journal]{IEEEtran}
\usepackage{amsmath}
\usepackage{footnote}
\usepackage{fancyhdr}
\usepackage{ragged2e}
\usepackage{enumitem}
\usepackage{amsfonts}
\usepackage{booktabs}
\usepackage{multirow}
\usepackage[small]{caption}
\usepackage{bbding}
\usepackage{pifont}
\usepackage{float}
\usepackage{subfigure}
\usepackage{graphicx}
\usepackage{url} 
\usepackage{algorithm, algorithmic}
\begin{document}

\title{TeKo: Text-Rich Graph Neural Networks \\with
External Knowledge}




\author{Zhizhi~Yu, Di~Jin, 
        Jianguo~Wei, Ziyang~Liu, Yue~Shang, Yun~Xiao, \\
        Jiawei~Han, 
        and~Lingfei~Wu
\IEEEcompsocitemizethanks{
\IEEEcompsocthanksitem Zhizhi Yu and Di Jin are with the College of Intelligence and Computing, Tianjin University, Tianjin 300350, China. 
E-mail: \{yuzhizhi, jindi\}@tju.edu.cn.
\IEEEcompsocthanksitem Jianguo Wei is with the College of Intelligence and Computing, Tianjin University, Tianjin 300350, China, and also with the Cumputer College, Qinghai Nationalities Unversity, Qinghai 810007, China.
E-mail: jianguo@tju.edu.cn.
\IEEEcompsocthanksitem Ziyang Liu is with the School of Software, Tsinghua University, Beijing 100080, China. 
E-mail: liu-zy21@mails.tsinghua.edu.cn.
\IEEEcompsocthanksitem Jiawei Han is with the Department of Computer Science, University of Illinois at Urbana-Champaign, Urbana, IL 61801. 
E-mail: hanj@cs.uiuc.edu.
\IEEEcompsocthanksitem Yue Shang, Yun Xiao and Lingfei Wu are with the JD.COM Silicon Valley Research Center, 675 E Middlefield Rd, Mountain View, CA 94043 USA. 
Email: \{yue.shang, xiaoyun1\}@jd.com,
lwu@email.wm.edu.
}
}

\maketitle

\begin{abstract}
Graph Neural Networks (GNNs) have gained great popularity in tackling various analytical tasks on  graph-structured data (i.e.,  networks).
Typical GNNs and their variants follow a message-passing manner that obtains network  representations by the feature propagation process along network topology, which however ignore the rich textual semantics (e.g., local word-sequence) that exist in many real-world networks. 
Existing methods for text-rich networks integrate textual semantics by mainly utilizing internal information such as topics or phrases/words, which often suffer from an inability to comprehensively mine the text semantics, limiting the reciprocal guidance between network structure and text semantics.
To address these problems, we propose a novel text-rich graph neural network with external knowledge (TeKo), in order to take full advantage of both structural and textual information within text-rich networks.
Specifically, we first present a
flexible heterogeneous semantic network that incorporates high-quality entities and interactions among documents and entities.
We then introduce two types of external knowledge, that is, structured triplets and unstructured entity descriptions, to gain a deeper insight into textual semantics. 
We further design a reciprocal convolutional mechanism for the constructed heterogeneous semantic network, enabling network structure and textual semantics to collaboratively enhance each other and learn high-level network representations. 
Extensive experimental results on four public text-rich networks as well as a large-scale e-commerce searching dataset illustrate the superior performance of TeKo
over state-of-the-art baselines.
\end{abstract}


\section{Introduction}
Networks are ubiquitous structures for abstracting and modeling relational data, such as social networks,  bibliographic networks and  biomedical networks. 
With their prevalence, it is particularly important to learn effective
representations of networks and apply them to downstream tasks.
Recently, graph neural networks (GNNs) \cite{GNNBook2022, DBLP:conf/icml/Ma0KW019, DBLP:journals/corr/LiTBZ15}, which exhibits significant power on
naturally capturing
both network structures and feature information, 
\begin{figure}[htp]
\centering
\subfigure[Classical GNNs following feature propagation along topology]{
\label{motivation_1}
\includegraphics[width=0.93\linewidth]{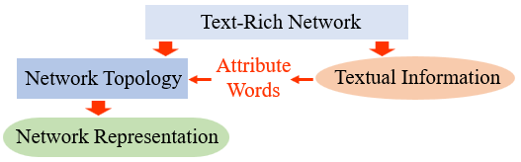}
}
\\
\subfigure[The proposed text-rich GNNs with  reciprocally propagation of topology and textual semantics guided by external knowledge]{
\label{motivation_2}
\includegraphics[width=0.93\linewidth]{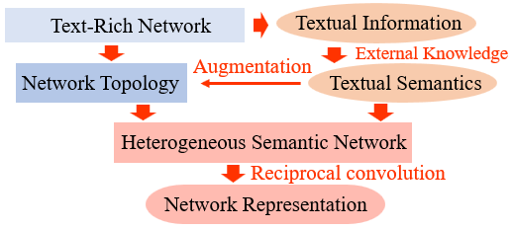}
}
\caption{\label{motivation} Comparison between classical GNNs and the proposed text-rich GNNs for text-rich network representations.}
\vspace{-0.5cm}
\end{figure}
have gained great success and been adapted in a wide range of application tasks, including
community detection \cite{DBLP:conf/ijcai/HeSJ0ZYZ20} and recommender system \cite{DBLP:conf/www/TanLZYZH20}.

The classical GNNs and their variants \cite{DBLP:conf/kdd/ZhangSHSC19,DBLP:conf/icml/Abu-El-HaijaPKA19}
follow a message-passing manner, where the most essential
part is the feature propagation along network topology. 
However, networks in the real-world usually contain nodes associated with rich textual information, which are called text-rich networks. 
Typical examples include
academic networks (e.g., DBLP) where document nodes are accompanied with their abstracts, and e-commerce network (e.g., Amazon) in which product nodes are attached with their descriptions. 
Under text-rich situation, existing GNNs may suffer from poor performance, since the propagation mechanism within node neighborhoods typically only treat textual information as attribute words, as shown
in Fig. \ref{motivation_1}, inevitably leading to the loss of some important semantic structure information (e.g., local word-sequence or global topic) contained in the text.

A few very recent studies \cite{DBLP:conf/wsdm/JinSYLZC021, DBLP:conf/icdm/YuJLHWT021}
have been dedicated to generalizing GNNs to text-rich networks. They incorporate semantic structures within the textual information, including local word-sequence and/or global topics, to original network structure for modeling text-rich networks. 
While doing so can leverage additional semantic information in text to a certain extent, these methods fail to well comprehend the semantic content of network data space, and reason over complex concepts and relational paths. This may introduce irrelevant information that negatively influence the performance of text-rich GNNs.
Our intuition is that, leveraging the knowledge provided from outside sources (e.g., Wikipedia and ConceptNet \cite{DBLP:conf/aaai/SpeerCH17}) can gain a deeper insight into the textual semantics (e.g., ``qos'' means ``quality of service'') and thus further facilitate the prediction of downstream tasks.

So an interesting yet important question is how to effectively leverage the external knowledge in order to design text-rich graph neural networks. There are two key challenges to consider.
First, how to identify appropriate and useful related knowledge in order to well comprehend the textual semantics underlying text-rich networks?
External knowledge usually consists of multiple types of data. For example, Wikipedia contains two types of data, i.e., structured triplets and unstructured entity descriptions.
Different data types represent different semantics, each of which reflects one aspect of textual information \cite{DBLP:conf/aaai/LinLSLZ15, DBLP:conf/coling/SunSQGHHZ20}. 
Therefore, it is important to comprehend the textual semantics via the reciprocal fusion of different data types of external knowledge.
Second, how to fully understand and leverage the acquired knowledge to facilitate the effective guidance of knowledge space to text-rich network data space?
As the structure and information contained in knowledge space and text-rich network data space are different, naively gluing information from these two spaces together may result in an over-complicated model
\cite{DBLP:conf/aaai/HuangXXDXLBXLY21}.
As a result, it is of great importance to design a more advanced model that can flexibly consider not only the information aggregation of each space, but also the information interaction among different spaces.
More importantly, the model itself should also correctly and adaptively learn the contribution of balance of network structure and textual information within text-rich networks aiming at given learning objectives.

In this paper, we focus on the above problems of semi-supervised learning on text-rich networks and propose a new \textbf{Te}xt-rich graph neural network with external \textbf{K}n\textbf{o}wledge, namely TeKo.
To this end, we first augment the text-rich network with a newly constructed heterogeneous document-entity network, as shown in Fig. \ref{motivation_2}, in order to incorporate entities and capture rich semantic structures among documents and entities.
We then introduce a knowledge-based entity representation module to adaptively extract useful external knowledge from structured triplets and unstructured entity descriptions.
In this way, the external knowledge is capable of helping comprehend the semantic content of network data space.
Finally, we design a discriminative propagation mechanism, based on reciprocal graph attention, in order to realize the interaction between network structure and textual semantics. 
During the intermediate training steps, these two parts are guided by each other and optimized collaboratively.
Meanwhile, by making full use of textual semantics, TeKo can also alleviate the topological limitations of GNNs \cite{DBLP:conf/nips/ZhuYZHAK20}  such as heterophily.

We summarize our main contributions as follows:
\begin{itemize}
\item 
To the best of our knowledge, we are the first to gain a deep insight into the textual semantics underlying text-rich networks through knowledge enhancement, empowering the effective fusion of both structural and textual information.
\item We present a novel text-rich graph neural network, TeKo, which innovatively employs the guidance of external knowledge space over network data space to learn text-rich network representations.
\item Extensive experiments demonstrate the superiority of the new approach TeKo over the state-of-the-art methods with significant improvements.
\end{itemize}

The rest of the paper is organized as follows. 
Section \ref{section2} gives the preliminaries. 
Section \ref{section3} proposes the new text-rich GNNs with external knowledge.
We conduct experiments in Section \ref{section4} and discuss related work in Section \ref{section5}.
Finally, we conclude in Section \ref{section6}.

\section{Preliminaries}
\label{section2}
We first introduce the terms and notations, and then give the problem definitions. We finally discuss GNNs which serve as the base of our proposed TeKo.

\subsection{Terms and Notations}
Given an undirected and unweighted text-rich network $G=(R, V, E)$, where $R$ is a set of document raw text, $V=\left\{v_{1}, \dots, v_{n}\right\}$ is a set of $n$ nodes, and $E = \{e_{ij}\} \subseteq V \times V$ is a set of edges.
The topological structure of $G$ is represented by an adjacency matrix $\bold{A} = [a_{ij}]\in \mathbb{R}^{n\times n}$, where $a_{ij} = 1$ if nodes $v_i$ and $v_j$ are connected, or $a_{ij}$ = 0 otherwise. 
The attribute matrix of $G$ is denoted as  
$\bold{X} \in \mathbb{R}^{n\times f}$, where attributes are extracted from document raw text $R$ and $f$ represents the number of attributes.
The notations we used throughout the
paper are summarized in Table \ref{tab:notations}.

\subsection{Problem Definitions}
We focus on two kinds of downstream tasks, namely, semi-supervised node classification and node clustering, to assess the learned text-rich network representations. 

\textbf{Semi-Supervised Node Classification}. Following the common semi-supervised learning setting, each node belongs to one out of $C$ classes and only a small part of nodes $V_L \ll n$ are associated with corresponding labels $Y_L$. 
\begin{table}[h!]
\begin{center}
 \caption{\label{tab:notations} Summary of notations.}
\scalebox{0.99}{
\begin{tabular}{@{}ll@{}}
	\toprule
	{\bf Notations}      & {\bf Descriptions}\\
	\midrule
	$G$ & A network. \\
	$R$ & The set of document raw text. \\
	$V, E$ & The sets of nodes and edges of a network. \\
	$e_{ij}$ & The edge between nodes $v_i$ and $v_j$.\\
    $a_{ij}$ & The connection between nodes $v_i$ and $v_j$.\\
	$\bold{A}, \bold{X}$ & The adjacency matrix and node attribute matrix.\\
    $\bold{D}$ & The node degree matrix. \\
    $\bold{x}_{i}$& The attribute vector of a given node $v_i$.\\
    \hline
    $(h, r, t)$ & A triplet in knowledge graph. \\
    $\bold{h}_i$ & The latent representation of a given node $v_i$.\\
   $\bold{e}_s$ & The triple representation of an entity node $w_i$.\\
    $\bold{e}_d$ & The textual representation of an entity node $w_i$.\\
   	$\sigma$ & The non-linear activation  function.\\
    $\mathcal{N}_i$ & The set of neighbors of node $v_i$.\\
    $\alpha$ & Weight of type-level attention.\\
    $\beta$ & Weight of node-level attention.\\
	\bottomrule
\end{tabular}
}
\label{notations}
\end{center}
\end{table}
\begin{figure*}
\setlength{\abovecaptionskip}{0.3cm}
    \centering
    \includegraphics[width=0.85\linewidth]{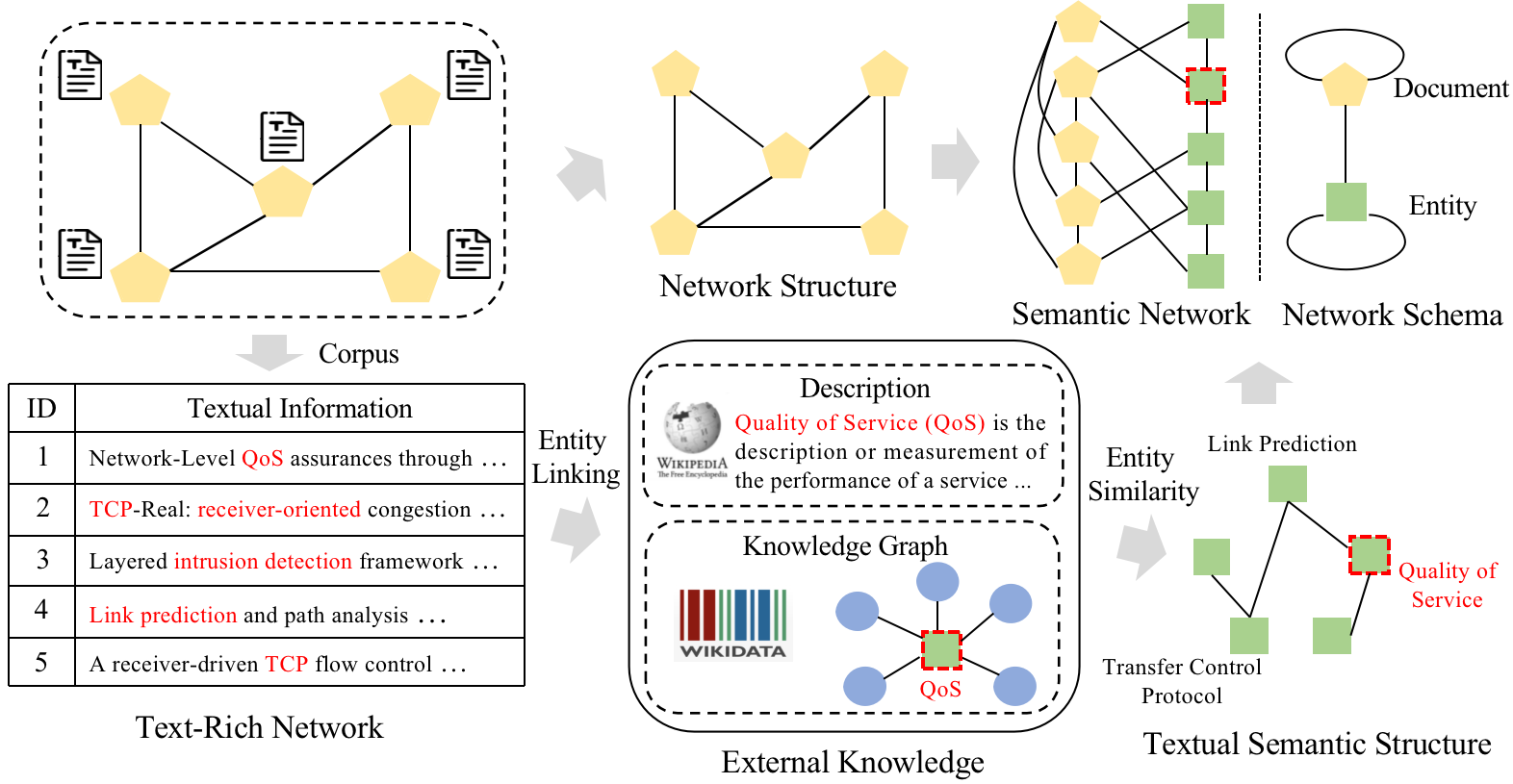}
    \caption{An illustrative example of a  text-rich network incorporating entities, where entities is annotated by TagMe \cite{DBLP:conf/cikm/FerraginaS10}.}
    \label{fig:framework}
\vspace{-0.4cm}
\end{figure*}
The objective of \emph{semi-supervised node classification} is to predict the labels of $V \verb|\| V_L$ by learning a function $\mathcal{F}$. 

\textbf{Node Clustering}. Following the unsupervised learning setting, that is, without any node labels. The goal of \emph{node clustering} is to design a mapping $\mathcal{F}$ to assign every node $v_i$ into one out of $C$ classes.

\subsection{Graph Neural Networks}
Graph Neural Networks (GNNs) \cite{DBLP:conf/iclr/KipfW17}
are a class of neural networks designed to  graph-related tasks in an end-to-end manner.
They typically learn node representations through an iterative aggregation of local network neighborhoods.
Formally, let $\bold{h}_i^{(k)}$ be the feature representation of node $v_i$ at the $k$-th layer, the message passing mechanism can be written as:
\begin{equation}
\begin{split}
&\bold{m}_i^{(k)}= \operatorname{MSG}^{(k)}({\bold{h}_i^{(k-1)}}),\\
&\bold{h}_i^{(k)} =  \operatorname{AGG}^{(k)}(
\bold{h}_i^{(k-1)}, \{\bold{m}_j^{(k)}: v_j \in \mathcal{N}(v_i)\})
\end{split}
\end{equation}
where $\bold{h}_i^{(0)}$ denotes the node’s attribute vector, $\bold{m}_i^{(k)}$ represents the message embedding, 
and $\mathcal{N}(v_i)$ is the local neighborhood of $v_i$.
GNNs work well on several network analytical tasks \cite{DBLP:conf/aaai/ZhuR0MLAK21}. 
But many networks in the real-world are text-rich. Since existing GNNs
typically treat text as attribute words alone, they will inevitably overlook important textual semantics. 
Therefore, it is of great significance to design a new text-rich GNN that takes full advantage of both
network structure and 
textual semantics.

\section{Methodology}
\label{section3}
We first give a brief overview of the proposed method, and then introduce three key components in detail.

\subsection{Overview}
\begin{figure*}
\setlength{\abovecaptionskip}{0.2cm} 
    \centering
    \includegraphics[width=0.90\linewidth]{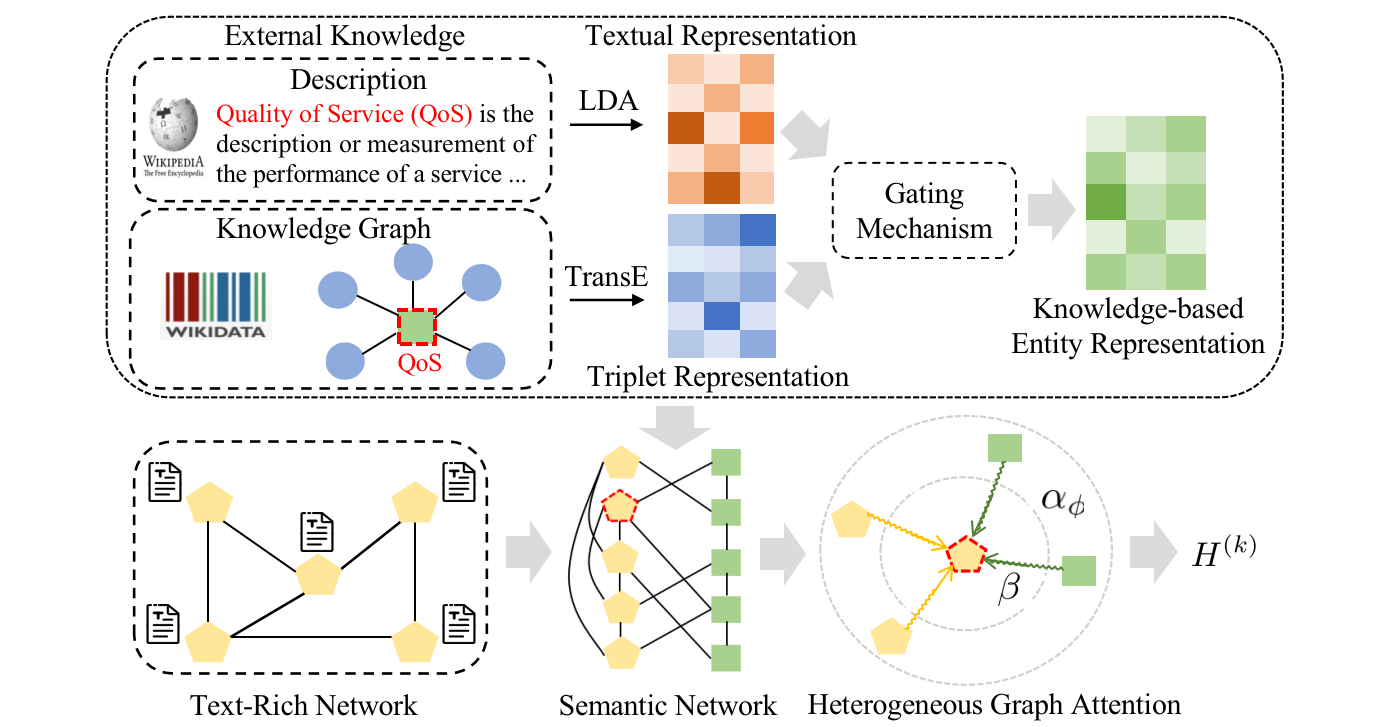}
    \caption{The architecture of TeKo which consists of three components, including semantic network generation, knowledge-based entity representation and heterogeneous graph attention.}
    \label{TeKo}
\vspace{-0.4cm}
\end{figure*}

To let the textual semantics  essentially provide supplementary information for text-rich network representations, we propose a novel text-rich graph neural network that can effectively integrate network structure and text semantics via the effective guidance of external knowledge space over network data space, namely TeKo.
The whole structure of the proposed approach is
illustrated in Fig. \ref{TeKo}, which consists of three main components:
semantic network generation, knowledge-based entity representation, and heterogeneous graph attention.
Specifically, for the semantic network generation, we construct a flexible heterogeneous document-entity framework for 
modeling the text-rich network, which 
entails both local and global semantic relationships among documents and entities.
For the knowledge-based entity representation, we extract useful and related knowledge from structured triples and unstructured entity descriptions, and accordingly learn jointly entity representations by automatically finding a balance
between these two types of information. 
This provides sufficient information for understanding text semantics and further enhance its integration with network structures.
For the heterogeneous graph attention, we introduce a discriminative propagation mechanism which adaptively aggregates the information from both text-rich network data space and knowledge space, enabling the model to realize a well balanced combination of network structure and textual semantics.

\subsection{Semantic Network Generation}
As an important auxiliary information of text-rich networks, textual semantics (e.g., local word-sequence) play an indispensable role in learning network representations.
To take full advantage of the underlying semantic structures within node textual information, we construct a heterogeneous semantic network that enables the integration of entities and captures the rich relationships among documents and entities. 
The semantic network generation consists of two parts: entity
network generation as well as whole network integration, as shown in Fig. \ref{fig:framework}.

\textbf{Entity Network Generation.}
As motivated, on a text-rich network, it is imperative to capture the underlying textual semantics. 
However, it will inevitably introduce a lot of useless information if we treat all words contained in the corpus as entities.
To this end, we consider recognizing high-quality entities from corpus
and map them to Wikipedia with the entity linking tool TagMe \cite{DBLP:conf/cikm/FerraginaS10}, that is, select entities above a predefined threshold $\delta_{tag}$.
After that, the edges among entities can be built according to the similarity of their initial knowledge-based representation (see Subsection 3.3 below). 
Actually, there are many ways to construct entity edges, and we uniformly choose the cosine similarity to generate edges between entities.

\emph{Cosine Similarity.} It uses the cosine value of the angle between two vectors to measure the entity similarity. Mathematically, given a text-rich network $G_D=(R, V_D, E_D)$, where each document node $v_{i}$ is associated with a textual description denoted as $d_{i}$. Let $V_W$ be the set of annotated entities, $\bold{e}_{i}$ be the knowledge-based representation of entity node $w_i$, then the similarity $s_{ij}$ between entity nodes $w_i$ and $w_j$ can be calculated as:
\begin{equation}
s_{ij} = \frac{\bold{e}_i\cdot \bold{e}_j}{|\bold{e}_i||\bold{e}_j|}.
\end{equation}

Then, the adjacency matrix $E_W$ can be obtained by choosing node pairs where the similarity is above a predefined threshold $\delta_{sim}$. Note that the constructed entity sub-network is static, where the entity nodes and edges do not change during the training process.

\textbf{Whole Network Integration.} 
We finally augment the original text-rich network into a heterogeneous semantic network, so as to explicitly capture both network topology and textual semantics. It includes two kinds of nodes (i.e., document nodes and entity nodes), and three kinds of edges (i.e., edges between document nodes from the original text-rich network representing paper relationships such as citations, edges between document nodes and entity nodes constructed based on the inclusion relationships between documents and entities, as well as edges between entity nodes capturing word-sequence semantics).
Formally, the heterogeneous semantic network is represented as:
\begin{equation}
G = (V_D \cup V_W, E_D \cup E_{DW} \cup E_W),
\end{equation}
where $E_{DW}$ is the set of edges between document nodes and entity nodes. In this way, by incorporating entities and relations, we can enrich the semantics of text-rich networks at a structure level.

\subsection{Knowledge-based Entity Representation}
To further comprehensively mine the textual semantics underlying text-rich networks and facilitate its interaction with text-rich network structure, we 
learn the entity representation using external knowledge. 
Considering that the information density of external knowledge is usually sparse and incomplete, we generate entity representation by encoding structured triplets and unstructured entity descriptions, respectively, and adaptively combine them with a learnable gating mechanism.

\textbf{Triplet Representation.}
Knowledge graph embedding is an effective way to parameterize
entities and relations as vector representations from structured triplets. 
Here we employ TransE \cite{DBLP:conf/nips/BordesUGWY13}, a simple and effective approach, to learn entity representations $\bold{e}_s\in \ \mathbb{R}^u$. Given a triplet $(h, r, t)$, let $\bold{r}$ be the representations of relation $r$, $\bold{h}$ and $\bold{t}$ be the representations of entities $h$ and $t$, respectively. 
TransE aims to embed each entity and relation by optimizing the translation principle $\bold{h} + \bold{r} \approx \bold{t}$, if $(h, r, t)$ holds. 
The score function is formulated as:
\begin{equation}
f(h, r, t) = - ||\bold{h} + \bold{r} - \bold{t}||^{2}_{2}, 
\end{equation}
where $\bold{h}$ and $\bold{t}$ are subject to the normalization constraint that the magnitude of each vector is 1. Intuitively, a large score of $f(h,r,t)$ indicates that the triplet is more likely to be a true fact in real-world, and vice versa. Notice that, we only consider triplets in which entity nodes within the heterogeneous semantic network are head (subject) instead of tail (object).

\textbf{Textual Representation.}
For each entity, we employ the paragraph of its corresponding Wikipedia page as the description.
As the description may represent an entity from various aspects, we adopt latent dirichlet allocation (LDA) \cite{DBLP:journals/jmlr/BleiNJ03}, a statistical model capable of modeling hidden topics behind texts, to learn the textual representation of each entity (denoted as $\bold{e_d} \in \mathbb{R}^u$). Specifically, LDA assumes that each description comes from a mixture of topics, where the topics are shared across the descriptions and the mixing proportion of each description is unique.
The generation process for each description can be iteratively divided into two steps: the first is to choose a topic with a certain probability, and the second is to select a word under this topic with a certain probability.

\textbf{Representation Fusion.}
Since both the structured triplets and unstructured entity descriptions
provide valuable information for an entity, we jointly integrate these two kinds of information for knowledge-based entity representation.
To achieve the optimal combination of triple representation $\bold{e}_s$ and textual representation $\bold{e}_d$, we introduce a learnable gating mechanism \cite{DBLP:conf/ijcai/XuQCH17} to determine how much the joint representation depends upon triples or description.  
Mathematically, given an entity node $w_i$, 
the joint representation $\bold{e}_i$ can be formulated as:
\begin{equation}
\bold{e}_i = \bold{g}_e \odot \bold{e}_s + (1-\bold{g}_e) \odot \bold{e}_d,
\end{equation} 
where $\bold{g}_e \in \mathbb{R}^u$ is a gating vector with elements in $[0,1]$ to balance the information from triples and description, and $\odot$ represents element-wise multiplication. 
Obviously, the joint representation with gate closer to 0 tends to use textual representation; whereas the joint representation with gate closer to 1 utilizes triple representation. More importantly, to constrain the value of each element in $[0, 1]$, we apply sigmoid function to calculate the gate $\bold{g}_e$:
\begin{equation}
\bold{g}_e = \operatorname{sigmoid}(\bold{\tilde{g}}_e),
\end{equation} 
where $\bold{\tilde{g}}_e$ is a real-value vector which is learned during the training process.

\subsection{Heterogeneous Graph Attention}
The \emph{core} of our approach is to
take full advantage of both network structure and textual semantics for text-rich network representations with the help of external knowledge.
For this purpose, we design a heterogeneous graph attention that performs information propagation on the augmented heterogeneous semantic network.
It considers not only the information aggregation in the text-rich network data space, but also the information guidance
of the knowledge space to the network data space.

Due to the heterogeneity of nodes in the augmented semantic network, different types of nodes have different feature spaces. 
Therefore, for a node $v_i$ with type $\phi_i$, we project its features into common space using a type-specific transformation matrix $\bold{W}_{\phi_i}$ as:
\begin{equation}
\bold{h}_i^{\prime} = \bold{W}_{\phi_i} \cdot \bold{h}_i,
\end{equation} 
where $\bold{h}_i^{\prime}$ is the projected feature of node $v_i$.

After that, to facilitate information propagation among neighboring nodes, we learn node representations from the perspective of network schema, which makes a heterogeneous network semi-structured, guiding the exploration of semantics of this network.
Typically, given a target node $v_i$, different types of neighboring nodes may have different impacts on it. 
Therefore, we employ type-level attention \cite{DBLP:conf/emnlp/HuYSJL19} to learn the importance of different types of neighboring nodes. 
To be specific, let $\bold{h_\phi}$ be the embedding of type $\phi$,
which is defined as the sum of the neighboring node embedding $\bold{h}_j^{\prime}$ with node $v_j \in \mathcal{N}_i$ under type $\phi$, that is:
\begin{equation}
\bold{h}_\phi = \sum\nolimits_{v_j} \hat{a}_{ij}\bold{h}_j^{\prime},
\end{equation} 
where $\bold{\hat{A}}=[\hat{a}_{ij}]=\bold{\tilde{D}}^{-\frac{1}{2}}\bold{\tilde {A}} \bold{\tilde {D}}^{-\frac{1}{2}}$ (${\bold{\tilde {A}} = \bold{A}+\bold{I}}$ stands for the adjacency matrix with self-loops, and $\bold{\tilde {D}}$ is degree matrix with $\tilde {d}_{ii} = \sum\nolimits_{v_j}\tilde {a}_{ij}$).

Then, based on the target node embedding $\bold{h}_i^{\prime}$ and the type embedding $\bold{h}_\phi$, the type-level attention weights can be calculated as:
\begin{equation}
\alpha_\phi =  \operatorname{softmax}_\phi(\operatorname{\sigma}(\eta_\phi^T [\bold{h}_i^{\prime}, \bold{h}_\phi])),
\end{equation}
where $\eta_\phi$ is the attention vector for the type $\phi$, $\sigma$ represents activation function such as LeakyReLU, and the softmax function is adopted to normalize across all the types.

On the other hand, considering that different neighboring nodes of the same type could also have different importance, we further apply node-level attention \cite{DBLP:conf/iclr/VelickovicCCRLB18} to learn the weights between nodes of the same type. 
Formally, given a target node $v_i$ with type $\phi$, let $v_j$ be its neighboring node with type $\phi'$, the node-level attention weights can then be computed as:
\begin{equation}
\beta_{ij} =  \operatorname{softmax}_{v_j}(\operatorname{\sigma}(\gamma^T \cdot \alpha_{\phi'} [\bold{h}_i^{\prime}, \bold{h}_j^{\prime}])),
\end{equation}
where $\gamma$ is the attention vector, and softmax is applied to normalize across all the neighboring nodes of the target node $v_i$.

By integrating the above process, the matrix form of the layer-wise propagation rule in heterogeneous graph attention can be defined as follows:
\begin{equation}
\bold{H}^{(k)} =  \operatorname{\sigma}(\sum\nolimits_{\phi \in \Phi} \bold{\mathcal{B}}_{\phi} \cdot \bold{H}_{\phi}^{(k-1)} \cdot \bold{W}_{\phi}^{(k-1)}),
\end{equation}
where $\Phi$ is the set of node types in the augmented semantic network. 
In this way, we can well realize reciprocal enhancement of information from both network structure and textual semantics with the guidance of external knowledge.

\subsection{Model Training}
After obtaining the final document representation, we can apply it to specific tasks and design different loss functions. 
For semi-supervised node classification, we 
define the loss function by using cross entropy as: 

\begin{equation}
\mathcal{L}=- \sum\limits_{ v_i\in V_L}\sum\limits_{c=1}^C \bold{y}_{i}[c]\operatorname{log}\bold{h}_{i}[c],
\end{equation}
where $V_L$ denotes the set of node indices that have labels, $C$ is the number of classes, and $\bold{y}_{i}$ is the one-hot label vector of node $v_i$.

For unsupervised learning, without any node labels, we define the loss function through negative sampling as:
\begin{equation}
 \mathcal{L}=- \sum\limits_{(v_i, v_j)\in \Omega}\operatorname{log}\sigma(\bold{h}^\top_i\cdot\bold{h}_j) - \sum\limits_{(v_m, v_n)\in \Omega^-}\operatorname{log}\sigma(-\bold{h}^\top_m\cdot\bold{h}_n),
\end{equation}
where $\sigma$ represents activation function, $\Omega$ is the set of positive node pairs, $\Omega^-$ is the set of negative node pairs (the complement of $\Omega$). 


\section{Experiments}
\label{section4}
We first introduce the experimental setup. We then compare the new approach TeKo with state-of-the-arts on three network analysis tasks, thereafter present an in-depth analysis of different components of TeKo and give the parameter analysis. We finally introduce the application on e-commerce search scenes.

\subsection{Experimental Setup}
\noindent\textbf{Datasets.} 
We conduct experiments on the following four public text-rich datasets, where
the basic information are summarized in Table \ref{tab:dataset}.

\begin{itemize}
\item \textbf{Cora-Enrich\footnote{http://zhang18f.myweb.cs.uwindsor.ca/datasets/}} is a text-rich
citation dataset composed of machine learning papers, where each node represents a document with textual description composed of its title and abstract. The documents are manually labeled with seven categories, e.g., \emph{Reinforcement Learning}, based on their academic topics.

\item \textbf{DBLP-Five}
\cite{DBLP:conf/kdd/TangZYLZS08} is extracted from the computer science bibliography website, where the textual information of each document contains its title and description.
According to the published domains, the documents
are labeled with five research areas, including \emph{High-Performance Computing, Software engineering, Computer networks, Theoretical computer science, and Computer graphics: Multimedia}.

\item \textbf{Hep-Small\footnote{https://www.cs.cornell.edu/projects/kddcup/datasets.html\label{data}} and Hep-Large}\textsuperscript{\ref {data}} are two citation datasets about scientific documents in physics, where 
Hep-Small contains 397 documents in three categories: \emph{Phys.Rev.Lett, Nucl.Phys.Proc.Suppl, and Commun.Math.Phys}; while Hep-Large contains 11,752 documents in four categories: \emph{Phys.Rev, Phys.Lett, Nucl.Phys, and JHEP}.
\end{itemize}

\begin{table}[h]
\begin{center}
\caption{The statistics of the datasets.}\label{tab:dataset}%
\begin{tabular}{@{}lllc@{}}
\toprule
 \textbf{Datasets} & \textbf{\#Nodes} & \textbf{\#Edges} & \textbf{\#Categories}\\
\midrule
\textbf{Hep-Small} & 397 & 812 & 3\\
\textbf{Cora-Enrich} & 2,708 & 5,429 & 7\\
\textbf{DBLP-Five} & 6,936 & 12,353 & 5\\
\textbf{Hep-Large} & 11,752 & 134,956 & 4\\
\bottomrule
\end{tabular}
\end{center}
\vspace{-0.3cm}
\end{table}

\begin{table*}[ht]
	\centering
	\caption{Node classification results with mean value and standard deviation in terms of Accuracy (\%) and Macro-F1 (\%). Bold and underline are used to show the best and the runner-up results.}
	\scalebox{1.0}{
        \begin{tabular}{c|cccccccc}
        		\toprule	\multirow{2}{*}{\textbf{Methods}} &	\multicolumn{2}{c}{\textbf{Hep-Small}} &\multicolumn{2}{c}{\textbf{Cora-Enrich}} & \multicolumn{2}{c}{\textbf{DBLP-Five}} & \multicolumn{2}{c}{\textbf{Hep-Large}}\\
        		\cmidrule{2-9}
             	& \textbf{Accuracy} & \textbf{Macro-F1} & \textbf{Accuracy} & \textbf{Macro-F1} & \textbf{Accuracy} & \textbf{Macro-F1}
             	& \textbf{Accuracy}
             	&\textbf{Macro-F1}\\
		\midrule
		GCN  & 61.54 $\pm$ 3.44 & 60.37 $\pm$ 4.17 & 88.71 $\pm$ 0.76 &  87.83 $\pm$ 0.90
		& 93.52 $\pm$ 0.47 & 93.19 $\pm$ 0.53
		& 50.24 $\pm$ 0.44 & 50.23 $\pm$ 0.45 \\
	GAT  & 64.87 $\pm$ 3.64 & 
	64.12 $\pm$ 4.37 & 88.96 $\pm$ 1.00 &  88.48 $\pm$ 1.11
		& 93.70 $\pm$ 0.34 & 93.30 $\pm$ 0.31
		& 50.17 $\pm$ 0.57 & 49.43 $\pm$ 1.00
		\\
		SGC  
		& 63.59 $\pm$ 3.94 & 63.01 $\pm$ 4.41
		& 88.00 $\pm$ 1.22 
		& 87.13 $\pm$ 0.97
		& 93.78 $\pm$ 0.46
		& 93.29 $\pm$ 0.55 & 47.69 $\pm$ 1.15 & 46.15 $\pm$ 1.31 \\
		DGI  & 
		59.72 $\pm$ 3.04 &
		59.30 $\pm$ 3.16 &
		79.57 $\pm$ 1.68 &
      	74.66 $\pm$ 1.62 & 
		83.48 $\pm$ 1.54 & 81.28 $\pm$ 1.64
		& 43.22 $\pm$ 0.80 & 39.28 $\pm$ 0.79 \\
		GMI  & 
		60.00 $\pm$ 2.98 &
		59.23 $\pm$ 2.96 &
		85.03 $\pm$ 0.46 &
        84.81 $\pm$ 0.45 & 
		92.29 $\pm$ 0.17 & 91.41 $\pm$ 0.20
		& 49.55 $\pm$ 0.24 & 49.44 $\pm$ 0.25 \\
		GraphSage  & 
		63.33 $\pm$ 2.31 & 62.07 $\pm$ 2.52 & 89.48 $\pm$ 0.81 & 88.38 $\pm$ 1.22 & 93.71 $\pm$ 0.70 & 93.35 $\pm$ 0.82 & 46.26 $\pm$ 0.85 & 45.54 $\pm$ 0.90\\
		AM-GCN  &
		\underline{68.21 $\pm$ 4.00} & 67.65 $\pm$ 4.36 & 89.04 $\pm$ 1.06 & 88.16 $\pm$ 1.43 & 93.52 $\pm$ 0.59 & 93.05 $\pm$ 0.65 & 47.17 $\pm$ 1.24 & 45.40 $\pm$ 1.51 \\
		Geom-GCN  &
		64.36 $\pm$ 2.42 & 64.33 $\pm$ 2.88 & 88.19 $\pm$ 1.65 & 87.09 $\pm$ 1.89 & \underline{94.57 $\pm$ 0.62} & \underline{94.19 $\pm$ 0.73} & 
		\underline{51.78 $\pm$ 0.80} & 
		\underline{51.42 $\pm$ 0.95}\\
		BiTe-GCN  & 
	67.44 $\pm$ 3.04 & 66.84 $\pm$ 3.06 & 89.59 $\pm$ 0.99 & 
	88.69 $\pm$ 0.74 & 93.81 $\pm$ 0.59 & 93.12 $\pm$ 0.58 & 50.43 $\pm$ 1.32 &
	50.12 $\pm$ 1.82 \\
	AS-GCN  &
		67.95 $\pm$ 4.48 & 
		\underline{68.06 $\pm$ 3.97} & 
		\underline{91.18 $\pm$ 1.52} & 
		\underline{90.29 $\pm$ 1.81} 
		& 94.17 $\pm$ 0.88 & 93.66 $\pm$ 0.97 & 
	50.92 $\pm$ 1.02  
	& 50.81 $\pm$ 0.88\\
	\midrule
    TeKo  
    & 
    \textbf{71.54 $\pm$ 3.88} & 
	\textbf{70.76 $\pm$ 4.37} & 
	\textbf{92.11 $\pm$ 0.86} &  
    \textbf{91.38 $\pm$ 0.95} & 
	 \textbf{95.12 $\pm$ 0.52} &  
	  \textbf{94.74 $\pm$ 0.68} &  \textbf{53.02 $\pm$ 1.03} &  \textbf{53.01 $\pm$ 1.01} \\
		\bottomrule
     \end{tabular}
            }
    \label{table 2}
    \vspace{-0.25cm}
\end{table*}

\noindent\textbf{Baselines.}
We evaluate the performance of our proposed TeKo by comparing it with ten state-of-art baselines. 

\begin{itemize}
\item \textbf{GCN} \cite{DBLP:conf/iclr/KipfW17} is a classical GNN which
derives node representations by defining convolutional operators on graph-structured data. 

\item \textbf{GAT} \cite{DBLP:conf/iclr/VelickovicCCRLB18} is an attention-based GNN which performs convolutional operations in the graph spatial domain and assigns different weights to neighbors.

\item \textbf{SGC} \cite{DBLP:conf/icml/WuSZFYW19} is a simplified GNN that reduces the complexity of model by removing nonlinearities and collapsing weight matrices between consecutive layers.

\item \textbf{DGI} \cite{DBLP:conf/iclr/VelickovicFHLBH19} is an unsupervised
GNN which maximizes
local mutual information via utilizing the graph’s patch representations.

\item \textbf{GMI} \cite{DBLP:conf/www/PengHLZRXH20}
is an unsupervised GNN that measures the correlation between input graphs and high-level representations through  graphical mutual information.

\item \textbf{GraphSage} \cite{DBLP:conf/nips/HamiltonYL17} is an inductive GNN leveraging sampler and aggregator to generate node  representations. 

\item \textbf{AM-GCN} \cite{DBLP:conf/kdd/0017ZB0SP20} is an adaptive
multi-channel GNN which extracts node representations via learning suitable weights to fuse the information from topology space and feature space. 

\item \textbf{Geom-GCN} \cite{DBLP:conf/iclr/PeiWCLY20} is a semi-supervised GNN designing a geometric aggregation mechanism to 
aggregate neighbor information of each node.

\item \textbf{BiTe-GCN} \cite{DBLP:conf/wsdm/JinSYLZC021} is a text-rich GNN that obtains node representations by integrating word sequence structure.

\item \textbf{AS-GCN} \cite{DBLP:conf/icdm/YuJLHWT021} is a text-rich GNN employing both local word-sequence and global topic semantic structures for node representations.

\end{itemize}

\noindent\textbf{Implementation Details.} 
For all baselines, we use the same parameter settings suggested by their papers. For BiTe-GCN and AS-GCN, we employ pre-trained 300-dimensional
GloVe embeddings \cite{DBLP:conf/emnlp/PenningtonSM14} to initialize entity representations. 
For our model, 
we use Wikipedia anchors to align mentions extracted from the textual description of each document node to Wikidata5M \cite{DBLP:journals/tacl/WangGZZLLT21}, which is a newly proposed large-scale knowledge graph containing 4M entities and 21M fact triplets. 
We set the layer number of  heterogeneous graph attention as 2, the Adam optimizer with an
initial learning rate of 0.005 and a weight decay of 5e-4. 
In addition, the threshold of TagMe and similarity score (cosine similarity) are searched in \{0.1, 0.2, 0.3, 0.4\} and \{0.5, 0.6, 0.7, 0.8, 0.9\}, respectively.
We set the activation function as LeakyReLU with slope
0.2, and apply a dropout rate of 0.5 to prevent overfitting. 
For a fair comparison of all methods, we 
generate 10 random
splits for training, validation and test. 

\subsection{Node Classification}
On the node classification task,  the goal is to predict the labels of the remaining nodes on the premise of giving a fraction of node labels. 
Since the variance of graph-structured data can be quite large, we report the average Accuracy and Macro-F1 
along with the standard deviation
of 10 independent
trials with different random seeds.

As presented in Table \ref{table 2}, we can find that TeKo performs consistently much better than all baselines across the four datasets.
Specifically, in terms of Accuracy, TeKo achieves up to 3.33\%, 0.93\%, 0.55\% and 1.24\% better accurate than the best baseline method on Hep-Small, Cora-Enrich, DBLP-Five and Hep-Large, respectively.
In terms of Macro-F1, TeKo is 2.70\%, 1.09\%, 0.55\% and 1.59\% more accurate than the best baseline method on these four datasets. 
These results not only demonstrate the superiority of comprehensively mining textual semantics underlying text-rich networks via external knowledge, but also validate the effectiveness of our new propagation mechanism that  facilitates the information interaction between network structure and textual information.
The performance of TeKo is much better than that of vanilla GCN (i.e., 10.00\%, 3.40\%, 1.60\% and 2.78\% relative improvements in Accuracy, and 10.39\%, 3.55\%, 1.55\%, and 2.78\% relative improvements in Macro-F1),
which implies that TeKo is capable of making a well balanced combination of both network structure and textual semantics within a text-rich network. 
Also of note, comparing with BiTe-GCN and AS-GCN which are also designed for text-rich networks, TeKo also achieves the best performance, which further verifies the significance of fully comprehending textual semantics for text-rich network representations with the guidance of external knowledge.

\subsection{Node Clustering}
We also conduct comparisons of these methods on node clustering. In this task, for each method,
the learned document embeddings are used as the input to K-Means algorithm, where $K$ is set to the number of clusters.
Since the performance of clustering is easily affected by the initial center,
we report the average normalized mutual information (NMI) and adjusted rand index (ARI) along with the standard deviation of 10 splits.

As shown in Table~\ref{Node Clustering(NMI)} and Table~\ref{Node Clustering(ARI)}, the proposed method TeKo performs the best across all the four datasets. To be specific, TeKo outperforms BiTe-GCN and AS-GCN, which also focus on extracting textual semantics within
text-rich networks, by 5.49\% and 3.07\% in terms of NMI, and 0.0506 and 0.0302 in terms of ARI (in the range of -1 to 1) on average on all the four networks. TeKo also performs better than the vanilla GCN by 6.52\% in terms of NMI and 0.064 in terms of ARI on overage. 
These results further validate the soundness of designing a text-rich graph neural network with external knowledge to take advantage 
of both structure and textual information within text-rich networks. Neither AM-GCN nor Geom-GCN is so competitive here. This may be mainly because they fail to fully utilize the semantics contained in the text, which significantly compromises their performance in the unsupervised clustering setting.

\begin{table}[h]
	\caption{Node clustering results with mean and standard deviation in terms of NMI (\%). Bold and underline are used to show the best and the runner-up results.}
    \resizebox{\linewidth}{!}{
      \begin{tabular}[width=\textwidth]
        {c|cccc}
        		\toprule
    	\textbf{Methods} &
	\textbf{Hep-Small}&
	\textbf{Cora-Enrich} & \textbf{DBLP-Five} & \textbf{Hep-Large}
        		\\
		\midrule
		GCN  & 24.37 $\pm$ 5.87 
		& 75.71 $\pm$ 1.30 
		& 81.07 $\pm$ 1.37
		& 12.18 $\pm$ 0.61 \\
	GAT  & 27.02 $\pm$ 7.82
	& 76.42 $\pm$ 1.68 
	& 81.46 $\pm$ 0.99 
	&  12.74 $\pm$ 0.44 \\
	SGC  
		& 25.33 $\pm$ 7.49
		& 74.77 $\pm$ 2.25
		& 82.10 $\pm$ 1.37 
		& 9.54 $\pm$ 0.85 \\
		DGI  & 
		20.55 $\pm$ 5.47 &
		62.99 $\pm$ 2.46 &
		79.57 $\pm$ 1.68 &
      	7.52 $\pm$ 0.69 \\
      	GMI  & 
		28.85 $\pm$ 8.34 &
		63.84 $\pm$ 2.89 &
		78.40 $\pm$ 0.37 &
      	11.90 $\pm$ 0.42 \\
		GraphSage  & 
		26.33 $\pm$ 6.29 & 
		77.46 $\pm$ 1.34 & 
		81.27 $\pm$ 1.76 & 
		7.98 $\pm$ 0.67 \\
		AM-GCN  &
		28.84 $\pm$ 5.57
		& 76.22 $\pm$ 2.23 
		& 80.90 $\pm$ 1.46 
		& 11.68 $\pm$ 0.79 \\
		Geom-GCN  &
		26.78 $\pm$ 5.43 
		& 74.70 $\pm$ 3.61 & 
		\underline{83.66 $\pm$ 1.63}
		& 13.62 $\pm$ 0.71 \\
		BiTe-GCN  & 
	27.25 $\pm$ 4.75 & 
	77.70 $\pm$ 1.86 & 
	81.92 $\pm$ 1.26 & 
	13.79 $\pm$ 0.63 \\
	AS-GCN  & 
	\underline{30.34 $\pm$ 7.39} & 
	\underline{79.89 $\pm$ 2.75} & 
	82.81 $\pm$ 2.20 & 
	\underline{14.09 $\pm$ 0.76} \\
	\midrule
    TeKo  
    & 
    \textbf{36.38 $\pm$ 7.92} & 
	\textbf{82.04 $\pm$ 1.88} & 
	\textbf{85.73 $\pm$ 1.46} &  
    \textbf{15.24 $\pm$ 1.00} \\
		\bottomrule
         \end{tabular}
    }
    \label{Node Clustering(NMI)}
\vspace{-0.25cm}
\end{table} 
\begin{table}[h]
	\caption{Node clustering results with mean and standard deviation in terms of ARI in the range of [-1, 1].}
    \resizebox{\linewidth}{!}{
      \begin{tabular}[width=\textwidth]
        {c|cccc}
        		\toprule
    	\textbf{Methods} &
	\textbf{Hep-Small}&
	\textbf{Cora-Enrich} & \textbf{DBLP-Five} & \textbf{Hep-Large}
        		\\
		\midrule
		GCN  & 0.1992 $\pm$ 0.0511 
		& 0.7631 $\pm$ 0.0180 
		& 0.8477 $\pm$ 0.0105
		& 0.1199 $\pm$ 0.0051 \\
	GAT  & 0.2205 $\pm$ 0.0714 
	& 0.7579 $\pm$ 0.0248
	& 0.8543 $\pm$ 0.0098 
	&  0.1243 $\pm$ 0.0050 \\
	SGC  
		& 0.2025 $\pm$ 0.0539
		& 0.7413 $\pm$ 0.0330
		& 0.8561 $\pm$ 0.0117 
		& 0.0983 $\pm$ 0.0082 \\
		DGI  & 
		0.1409 $\pm$ 0.0410 &
		0.6001 $\pm$ 0.0388 &
		0.6570 $\pm$ 0.0306 &
      	0.0740 $\pm$ 0.0068 \\
      	GMI  & 
		0.2414 $\pm$ 0.0735 &
		0.6048 $\pm$ 0.0376 &
		0.8225 $\pm$ 0.0333 &
      	0.0788 $\pm$ 0.0021 \\
		GraphSage  & 
		0.2098 $\pm$ 0.0491 & 
		0.7805 $\pm$ 0.0221 & 
		0.8530 $\pm$ 0.0138 & 
		0.0801 $\pm$ 0.0073 \\
		AM-GCN  &
		\underline{0.2615 $\pm$ 0.0755}
		& 0.7659 $\pm$ 0.0254 
		& 0.8506 $\pm$ 0.0133 
		& 0.1127 $\pm$ 0.0117 \\
		Geom-GCN  &
		0.2016 $\pm$ 0.0407
		& 0.7535 $\pm$ 0.0342 & 
		\underline{0.8743 $\pm$ 0.0133}
		& 0.1344 $\pm$ 0.0084 \\
		BiTe-GCN  & 
	0.2312 $\pm$ 0.0493 & 
	0.7744 $\pm$ 0.0279 & 
	0.8532 $\pm$ 0.0143 & 
	0.1346 $\pm$ 0.0098 \\
		AS-GCN  & 
	0.2564 $\pm$ 0.0794 & 
	\underline{0.8073 $\pm$ 0.0248} & 
	0.8656 $\pm$ 0.0190 & 
	\underline{0.1360 $\pm$ 0.0107}\\
	\midrule
    TeKo   
    & 
    \textbf{0.3215 $\pm$ 0.0678} & 
	\textbf{0.8315 $\pm$ 0.0251} & 
	\textbf{0.8837 $\pm$ 0.0148} & 
    \textbf{0.1491 $\pm$ 0.0132} \\ 
		\bottomrule
         \end{tabular}
    }
    \label{Node Clustering(ARI)}
\vspace{-0.4cm}
\end{table} 
\begin{figure*}[htp]
\centering
\subfigure[GCN]{
\label{}
\includegraphics[width=0.18\linewidth]{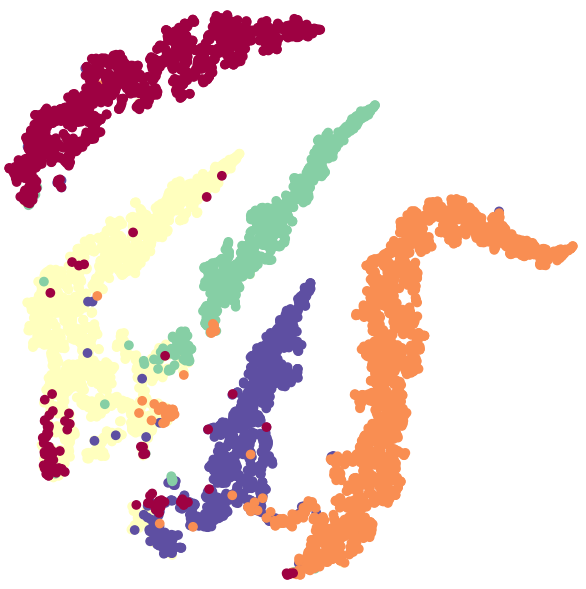}
}
\subfigure[GAT]{
\includegraphics[width=0.18\linewidth]{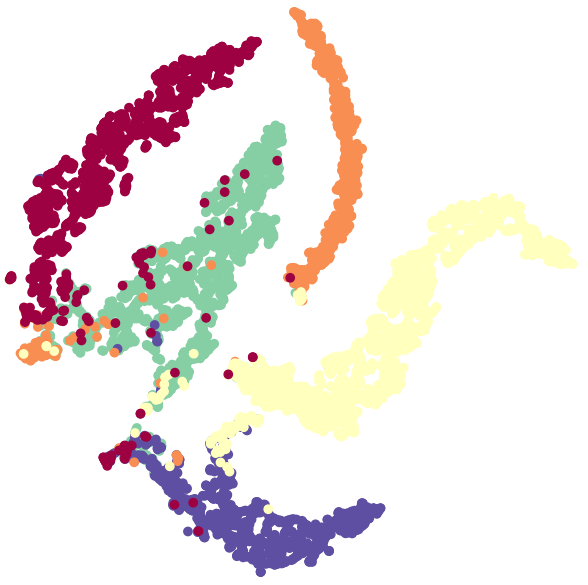}
}
\subfigure[Geom-GCN]{
\includegraphics[width=0.18\linewidth]{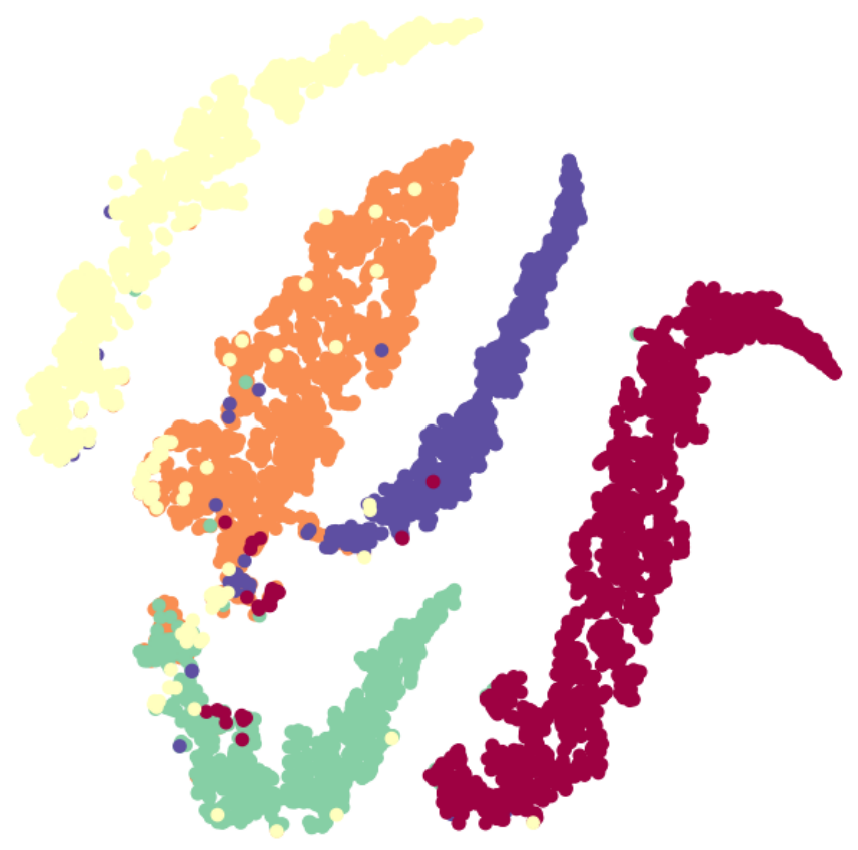}
}
\subfigure[AS-GCN]{
\includegraphics[width=0.18\linewidth]{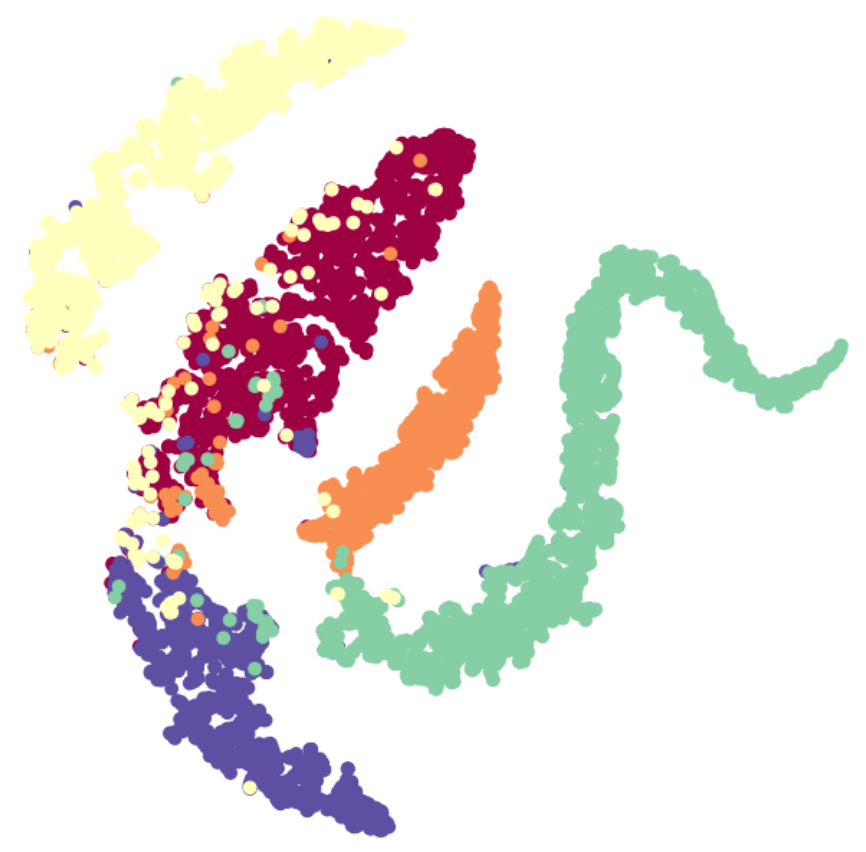}
}
\subfigure[TeKo]{
\includegraphics[width=0.18\linewidth]{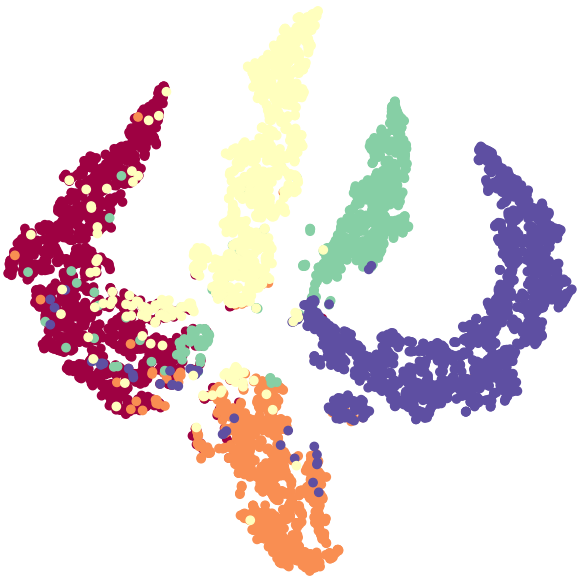}
}
\caption{\label{tsne}The visualization of the node representations learned by (a) GCN, (b) GAT, (c) Geom-GCN, (d) AS-GCN and (e) TeKo on the DBLP dataset. Different colors correspond to different categorical labels in ground truth.}
\end{figure*}
\begin{table*}[htp]
	\centering
	\caption{Comparisons of our TeKo with its three variants on node classification in terms of Accuracy (\%) and Macro-F1 (\%).}
	\scalebox{0.97}{
        \begin{tabular}{l|cccccccc}
        		\toprule
        		\multirow{2}{*}{\textbf{Methods}} &	\multicolumn{2}{c}{\textbf{Hep-Small}} &\multicolumn{2}{c}{\textbf{Cora-Enrich}} & \multicolumn{2}{c}{\textbf{DBLP-Five}} & \multicolumn{2}{c}{\textbf{Hep-Large}}\\
        		\cmidrule{2-9}
             	& \textbf{Accuracy} & \textbf{Macro-F1} & \textbf{Accuracy} & \textbf{Macro-F1} & \textbf{Accuracy} & \textbf{Macro-F1}
             	& \textbf{Accuracy}
             	&\textbf{Macro-F1}\\
		\midrule
    TeKo  & 
    71.54 $\pm$ 3.88 & 
	70.76 $\pm$ 4.37 & 
	\textbf{92.11 $\pm$ 0.86} &  
    \textbf{91.38 $\pm$ 0.95} & 
	 \textbf{95.12 $\pm$ 0.52} &  
	  \textbf{94.74 $\pm$ 0.68} &  \textbf{53.02 $\pm$ 1.03} &  \textbf{53.01 $\pm$ 1.01} \\
		- w/o Triplet & 
		69.49 $\pm$ 3.33 
		& 69.06 $\pm$ 3.30
		& 91.18 $\pm$ 1.89 
		& 90.86 $\pm$ 1.97
		& 94.90 $\pm$ 0.64 
		& 94.54 $\pm$ 0.71 
		& 50.66 $\pm$ 1.19 
		& 49.83 $\pm$ 1.42 \\
	- w/o Textual  &  
	\textbf{73.33 $\pm$ 3.48}
	& \textbf{72.94 $\pm$ 4.03}
	& 90.30 $\pm$ 1.58
	& 89.48 $\pm$ 1.60
	& 94.99 $\pm$ 0.46
	& 94.66 $\pm$ 0.54
	& 50.94 $\pm$ 0.95
	& 50.07 $\pm$ 0.93
		\\
		TeKo (concatenation)
		& 66.92 $\pm$ 2.68 
		& 66.63 $\pm$ 2.90
		& 90.89 $\pm$ 1.36
		& 90.18 $\pm$ 1.82
		& 94.24 $\pm$ 0.47
		& 93.78 $\pm$ 0.42
		& 50.98 $\pm$ 1.01
		& 50.37 $\pm$ 1.40  \\
		\bottomrule
     \end{tabular}
    }
    \label{Ablation Study}
\vspace{-0.4cm}
\end{table*}


\subsection{Visualization}
To provide a more intuitive comparison, we conduct embedding visualization of GCN, GAT, Geom-GCN, AS-GCN and our proposed TeKo on DBLP dataset as an example.
We use t-SNE \cite{29} to downscale the
learned node representations to a two-dimensional space, where different colors mean different labels. 
Therefore, a desirable visualization result refers to that nodes belonging to the same category (in the same color) should be close to each other.

From Fig. \ref{tsne}, we can find that the visualization results of GCN and GAT are not so satisfactory,
as nodes with the
same color are dispersed and nodes with different colors are mixed with each other. 
The results of Geom-GCN
and AS-GCN are relatively better but the borders between different classes are not so clear. 
Apparently, the visualization of TeKo performs better, where the learned representations have a higher intra-class similarity and
form more discernible clusters. 

\subsection{Ablation Study}
Similar to most deep learning models, our proposed method TeKo also contains some important components that may have a significant impact on the performance. 
To test the effectiveness of each component, we conduct experiments on comparing it with three variations. The variants are as follows: 
1) TeKo of removing triplet representation of entity, named as w/o triplet, 
2) TeKo of removing the textual representation of entity, named as w/o textual,
and 3) TeKo of employing concatenation operator instead of gating mechanism to generate the knowledge-based entity representation, named as TeKo (concatenation).

We take their comparison on node classification in terms of Accuracy and Macro-F1 as an example.
From the results in Table \ref{Ablation Study}, we can draw the following conclusions: 
(1) The results of TeKo are better than that of
its three variants in most cases (except on Hep-Small), indicating the effectiveness of
using both structured triplets and unstructured entity descriptions for entity representation together. 
(2) TeKo w/o textual is in general better than TeKo w/o triplet on three out of the four datasets, which implies the textual representation of entity plays a more vital role for fully comprehend textual semantics and promote its interaction with network structure.
(3) Compared to Teko w/o gating that uses concatenation operator to fuse the triple representation and textual representation of an entity, the improvement
brought by Teko is more significant, which illustrates the rationality of adaptively fusing these two representations aiming to the given learning
objectives.

\begin{figure*}[htbp]
\vspace{-0.08cm}
\centering
\subfigure[Hep-Small]{\includegraphics[width=.23\textwidth]{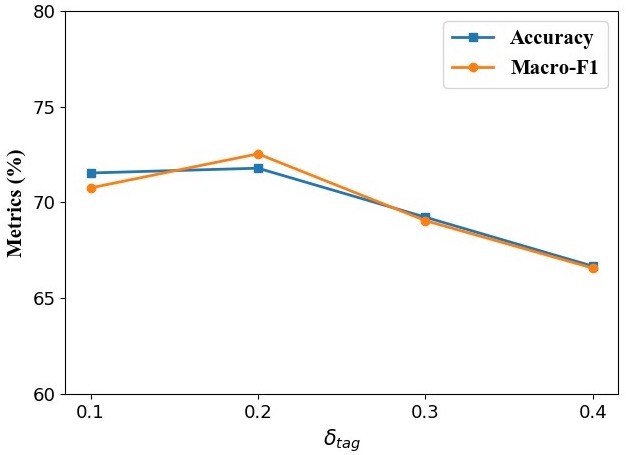}}
\subfigure[Cora-Enrich]{\includegraphics[width=.23\textwidth]{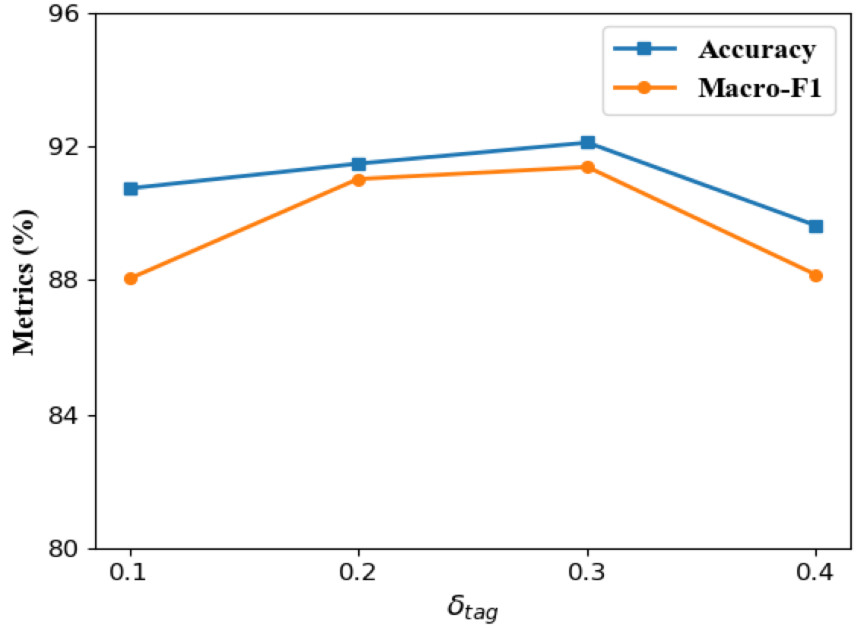}}
\subfigure[DBLP-Five]{\includegraphics[width=.23\textwidth]{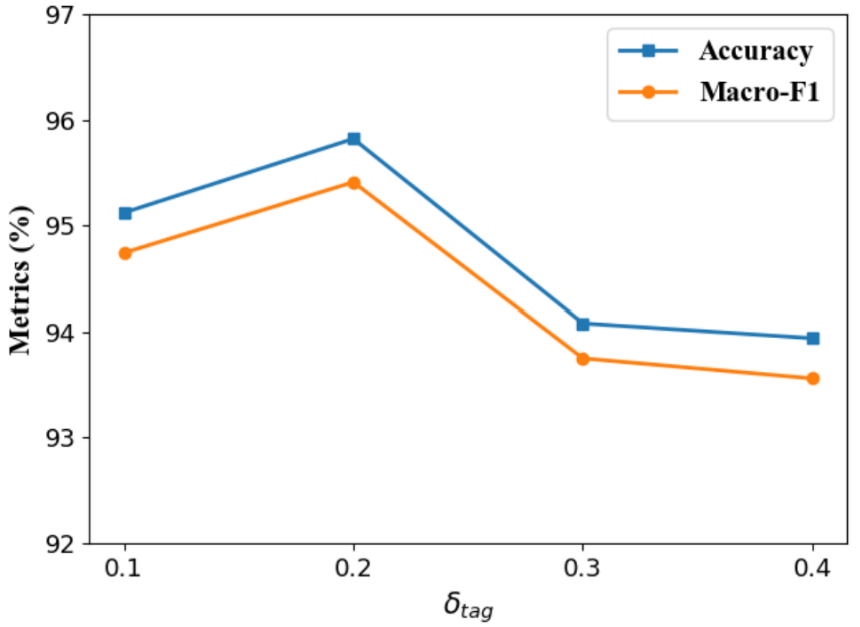}}
\subfigure[Hep-Large]{\includegraphics[width=.23\textwidth]{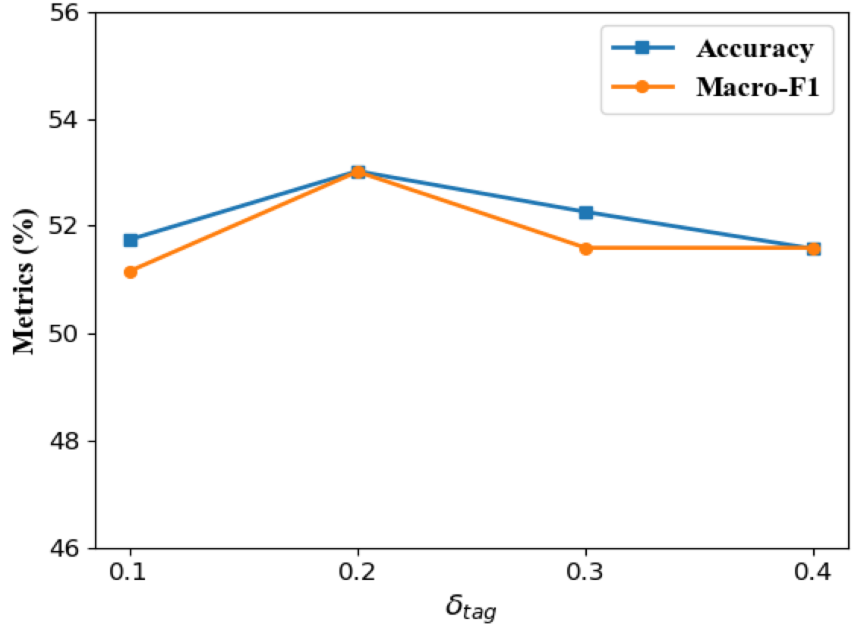}}
\caption{Analysis results of the threshold $\delta_{tag}$.}
\label{parameter_tag}
\vspace{-0.2cm}
\end{figure*}
\begin{figure*}[htbp]
\centering
\subfigure[Hep-Small]{\includegraphics[width=.23\textwidth]{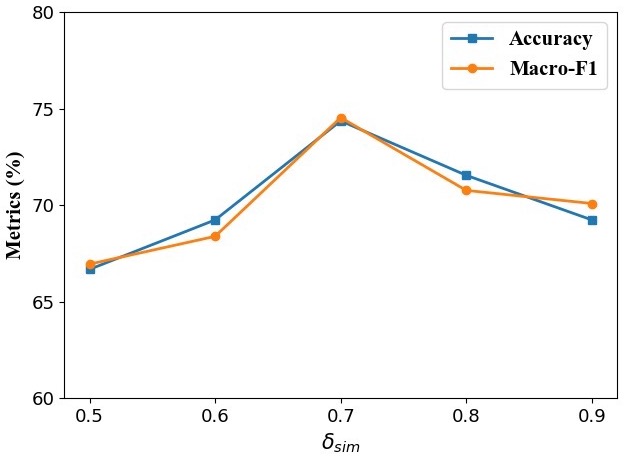}}
\subfigure[Cora-Enrich]{\includegraphics[width=.23\textwidth]{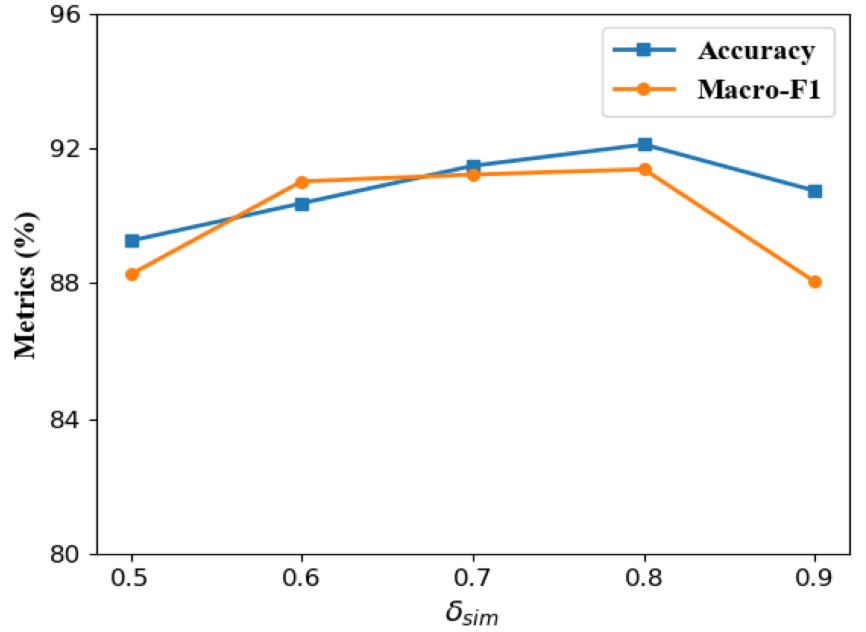}}
\subfigure[DBLP-Five]{\includegraphics[width=.23\textwidth]{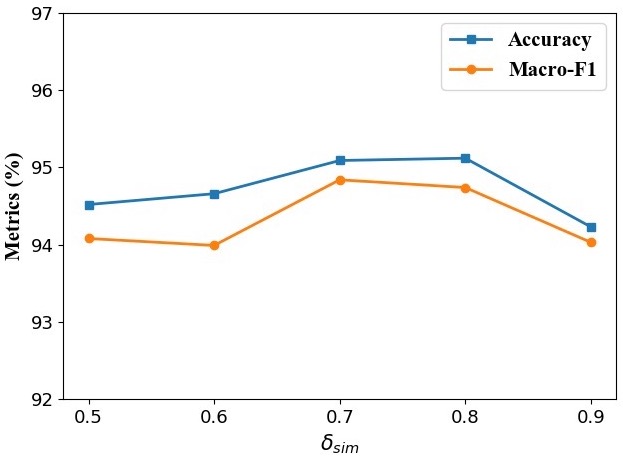}}
\subfigure[Hep-Large]{\includegraphics[width=.23\textwidth]{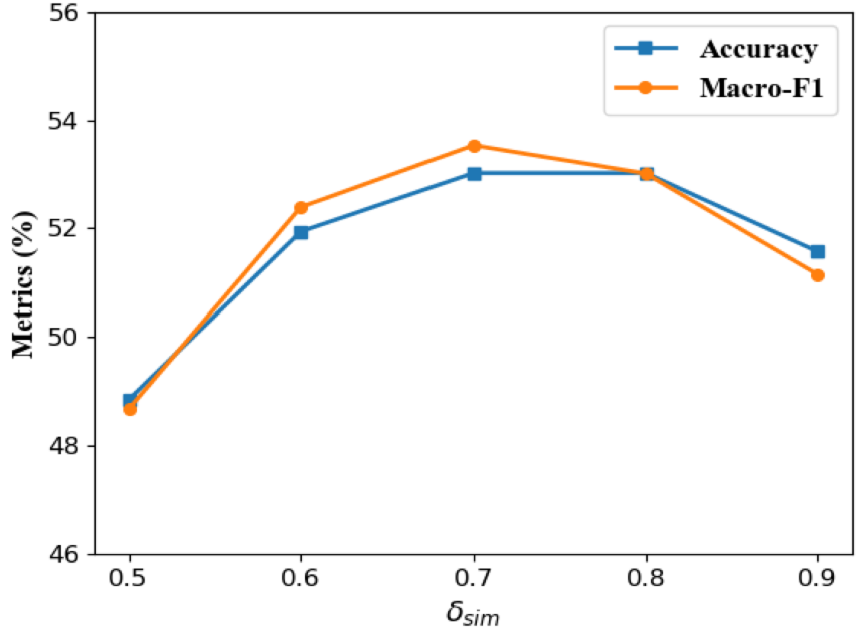}}
\caption{Analysis results of the threshold $\delta_{sim}$.}
\label{parameter_sim}
\vspace{-0.4cm}
\end{figure*}
\subsection{Parameter Analysis}
We investigate the sensitivity of two main parameters: the threshold $\delta_{tag}$ of TagMe and the threshold $\delta_{sim}$ of entity edge similarity. We take the node classification on all the four text-rich datasets as an example
and report the Accuracy and Macro-F1 values.

\textbf{Analysis of $\delta_{tag}$.}
The threshold $\delta_{tag}$ determines the number of entities within the generated semantic network. We vary its value from 0.1 to 0.4 and the corresponding results
are shown in Fig. \ref{parameter_tag}. With the increase of this threshold $\delta_{tag}$, the performance shows a trend of first rising and then descending.
It is reasonable since a too small threshold of TagMe would introduce some noise entities, whereas a too large threshold may filter some informative entities and thus weaken the reciprocal guidance between text semantics and network structure.

\textbf{Analysis of $\delta_{sim}$.}
The threshold $\delta_{sim}$ determines the number of edges between entities, which represents the local word-sequence semantic structure underlying the corpus.
We vary its value from 0.5 to 0.9 and the corresponding results are shown in Fig. \ref{parameter_sim}.
With the increase of this threshold of entity edge similarity, the values of metrics, including Accuracy and Macro-F1, also increase first and then start to descend. 
This is probably because a small number of edges between entities would result in information loss and ineffective information propagation, whereas too many edges between entities would introduce more noise.

\subsection{Application on  E-commerce Search}
To further verify the effectiveness of our proposed new approach, we collect an e-commerce searching dataset and apply TeKo on it to solve the problem of relevance matching, that is,  predicting whether the current example pair (query and item) is relevant or not. 
The collected dataset contains million-scale queries or items, and their detailed statistics are shown in Table~\ref{JD-datasets}.

\begin{table}[h]
\begin{center}
\begin{minipage}{200pt}
\caption{The statistics of the e-commerce  dataset.}\label{JD-datasets}
\begin{tabular}{@{}ccccc@{}}
\toprule
 \textbf{Datasets} 
 & \textbf{\#Queries} & \textbf{\#Items} &
\textbf{\#Edges} \\
\midrule
\textbf{Training set} &3,284,480 &1,307,557 &7,525,355 \\
\textbf{Validation set} &3,108 &30,097 &30,661 \\
\bottomrule
\end{tabular}
\end{minipage}
\end{center}
\vspace{-0.4cm}
\end{table}

\begin{figure}[h]
    \centering
    \includegraphics[width=0.35\textwidth]{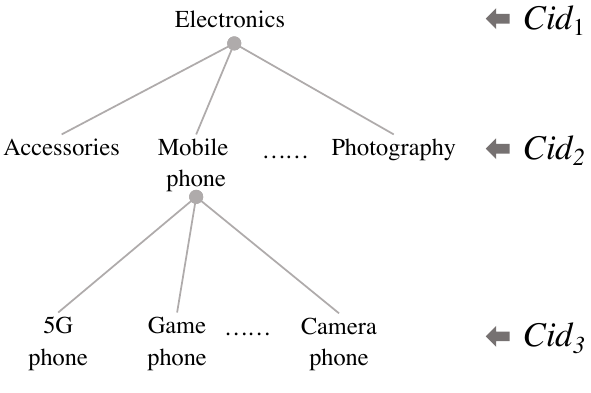}
    \caption{An illustrative example of tree-shape category information in our e-commerce search.}
    \label{fig:category}
\vspace{-0.2cm}
\end{figure}

\textbf{External Knowledge in E-commerce Search.} We use the category information to serve as the external knowledge in e-commerce search. Specifically, the category information includes three different levels of categories, i.e., $Cid_{1}$, $Cid_{2}$, and $Cid_{3}$. They present as a tree-shape structure as shown in Fig.~\ref{fig:category}. Adding this information is helpful to analyze the semantics of the current query or items. For example, for an item whose title is ``red mac 2020 (made in US)'', we cannot ensure whether is an electronic product or a cosmetic product. But if we further know that its $Cid_{1}$ is ``Clothes \& Cosmetics'', then we can easily infer that this is a lipstick.

\textbf{Baselines.} Since this relevance matching
problem lies in the field of natural language matching, we compare our new approach TeKo with seven state-of-arts in this topic.
\begin{itemize}
    \item \textbf{MV-LSTM} \cite{mv-lstm} is a deep model which assigns the importance score of each local keyword using rich context information to match two sentences.
    \item \textbf{K-NRM} \cite{k-nrm} is a kernel-based semantic matching model that uses embedding layer, translation, and kernel pooling to capture the word-level interaction relationship between both sentences.
    \item \textbf{ARC-I} \cite{arc} matches two sentences with their representations by using a multi-layer perceptron (MLP), which is essentially the Siamese architecture.
    \item \textbf{ARC-II} \cite{arc} is an advanced version of ARC-I. Compared to ARC-I, ARC-II more focuses on the interaction relationship between two sentences.
    \item \textbf{MatchPyramid} \cite{matchpy} employs hierarchical convolution to capture different-level matching patterns contained in both sentences, such as unigram, n-gram and n-term. 
    \item \textbf{DUET} \cite{duet}
    calculates the final relevance score by  summing the scores from the representation-based embedding and the interaction-based embedding. 
    \item \textbf{BERT2DNN}
     \cite{DBLP:conf/icdm/JiangSLSXXXYJ20} is a data-driven model which adopts the techniques of 
     transfer learning and knowledge distillation for search relevance.
\end{itemize}

\textbf{Metrics.} We choose six evaluation metrics to measure the model’s quality, including Area Under receiver operator characteristic Curve (AUC), Accuracy, Precision, F1-score, Recall, and  False Negative Rate (FNR). The lower FNR implies the better model, while the other metrics are opposite.
In particular, AUC is the most important of these metrics.

\textbf{Experimental Results and Analysis.} From the results in Table~\ref{JD-results}, we can find that our TeKo consistently outperforms all the baseline methods.
Especially, compared to the popular e-commerce search algorithm BERT2DNN, which uses transfer learning
and 
\begin{table}[h]
\begin{center}
\caption{Comparisons on an e-commerce searching dataset, where (*) denotes the dominant metric. The lower FNR implies the better model, while the other metrics are opposite.}\label{JD-results}%
\resizebox{\linewidth}{!}{
        \begin{tabular}[width=0.7\linewidth]{@{}c|cccccc@{}}
\toprule
 \textbf{Methods} &
 \textbf{AUC(*)} &
 \textbf{Accuracy} &
 \textbf{Precision} &
 \textbf{F1-score} & \textbf{Recall} & \textbf{FNR} \\
\midrule
MV-LSTM & 0.8760 & 0.8069 & 0.8392 & 0.7256 & 0.7541 &0.2459 \\
ARC-I & 0.7945 & 0.7392 & 0.8005 & 0.6329 & 0.6830 & 0.3170 \\
K-NRM & 0.8424 & 0.7899 & 0.7341 & 0.6919 & 0.7333 & 0.2667 \\
ARC-II & 0.8466 & 0.7913 & 0.6931 & 0.7002 &0.7439 & 0.2561 \\
MatchPyramid & 0.8758 & 0.8278 & 0.7741& 0.7451 & 0.7683 & 0.2317  \\
DUET & 0.8682 & 0.8156 & 0.8159 & 0.7253 & 0.7337 & 0.2663 \\
BERT2DNN & 0.8906 & 0.7896 & 0.8565 & 0.8362 & 0.8169 & 0.1831 \\
\midrule
TeKo & \textbf{0.9092} & \textbf{0.8319} & \textbf{0.8648} & \textbf{0.8735} & \textbf{0.8824} & \textbf{0.1176} \\
\bottomrule
\end{tabular}
}
\end{center}
\vspace{-0.5cm}
\end{table}
knowledge distillation to improve result relevance, 
the superiority of TeKo is even up to 4.23\% and 3.73\% improvements in Accuracy and F1-score, respectively.
These results not only demonstrate the effectiveness of our TeKo in capturing query-item pair correlation, but also shows the rationality of our adopting a more advanced way (i.e., utilize external knowledge to comprehend text semantics underlying text-rich situations for high-level relevance matching).

\section{Related Work}
\label{section5}
In line with the focus of our work, we briefly review some related studies, including classical graph neural networks, text-rich graph neural networks,  knowledge-enhanced graph neural networks, and graph neural networks for text analysis.

\textbf{Classical Graph Neural Networks.} 
Graph neural networks (GNNs) have attracted considerable research interests due to the ability to model graph-structured data. 
For example, Bruna \emph{et al}. \cite{DBLP:journals/corr/BrunaZSL13}
first designs the graph convolutional operation in Fourier domain through the graph Laplacian. 
Defferrard \emph{et al}. \cite{1} futher promote the efficiency using the Chebyshev polynomial expansion. 
After that comes GCN \cite{DBLP:conf/iclr/KipfW17}, a simplified graph convolutional operation that aggregates information from nodes' one-hop neighbors. 
GraphSAGE \cite{DBLP:conf/nips/HamiltonYL17} employs the mean/max/LSTM pooling to sample and aggregate the information from neighbors.
GAT \cite{DBLP:conf/iclr/VelickovicCCRLB18} assigns different weights to neighbors with a node-level attention mechanism.
Though these methods can be used for text-rich network representations, they suffer from an inability to fully consider and mine the text semantics (e.g., local word-sequence or global topic) underlying text-rich networks, which is particularly important for information propagation along network topology. 

\textbf{Text-Rich Graph Neural Networks.} 
Recently, many attentions have been paid to text-rich network representations. 
For instance, HyperMine \cite{DBLP:conf/cikm/ShiSLZHLZW0019}
is designed for discovering hypernymy from text-rich networks.
NetTaxo \cite{DBLP:conf/www/ShangZLL020} focuses on topic taxonomy construction, and integrates text data and network structures simultaneously.
LTRN \cite{DBLP:conf/www/ZhangZDS021} designs a minimally-supervised categorization framework based on personalized PageRank-based
neighborhood sampling and attentive aggregation from a text-rich network prospective. 
BiTe-GCN \cite{DBLP:conf/wsdm/JinSYLZC021} uses bidirectional convolution of topology and features to depict original network structure and local word-sequence structure extracted from text, and on this basis, 
AS-GCN \cite{DBLP:conf/icdm/YuJLHWT021} takes the global topic semantic structure into account. 
However, the above methods
can not well comprehend the semantic content contained in text-rich network, and reason over complex concepts and relational paths, limiting the reciprocal guidance between network structure and text semantics.


\textbf{Knowledge-Enhanced Graph Neural Networks.} 
As GNNs become the most eye-catching tools for node representations, several efforts have been made to combine external knowledge and GNNs to boost the performance of downstream tasks. 
For instance,
KGCN \cite{DBLP:conf/www/WangZXLG19} captures users’ preferences on the knowledge graph for recommender systems.
KGAT \cite{DBLP:conf/kdd/Wang00LC19} presents a knowledge-aware recommendation
by explicitly modeling the high-order connectivity with semantic relations 
in collaborative knowledge graph.
More recently, 
RGHAT \cite{DBLP:conf/aaai/ZhangZZ0XH20} designs a two-level attention mechanism which uses
the inherent and valuable neighborhood information surrounding an entity for knowledge graph completion.
Caps-GNN \cite{DBLP:conf/cikm/LiLZHWYW20}
proposes a novel knowledge-enhanced personalized review generation model that adopts capsule graph neural networks and knowledge graph to capture both aspect- and word-level user preference.
CompareNet \cite{DBLP:conf/acl/HuYZZTSD020} utilizes topics and entities extracted from the text to enrich news representation, then compares news to the external knowledge through entities to detect fake news. 
However, how to utilize the guidance of external knowledge to facilitate text-rich network representations is still an area that needs to be explored urgently. 

\textbf{Graph Neural Networks for Text Analysis.} 
Another research line relevant to our work is to apply GNNs for text analysis.
For example, Text GCN \cite{DBLP:conf/aaai/YaoM019} turns text classification into a node classification problem by constructing a heterogeneous word document graph for a whole corpus.
HGAT \cite{DBLP:conf/emnlp/HuYSJL19} enriches the semantics of the short texts via incorporating the topics, entities and the relations.
TensorGCN \cite{DBLP:conf/aaai/LiuYZWL20} constructs a
text graph tensor to describe contextual
information within text, and further integrate these information through 
generalizing the GNN into a tensor version.
TextING \cite{DBLP:conf/acl/ZhangYCWWW20}
treats each text as an individual graph
and learns word interactions at the text level for inductive text classification.
TextGTL \cite{DBLP:conf/ijcai/LiPPLW21} proposes a non-heterogeneous graph construction method which jointly considers
different linguistic information, including semantics, syntax and sequence context for text classification. In summary, the above GNNs mainly focus on establishing  associations for independent texts via building a graph
structure, which essentially lies in the field of text analysis, rather than text-rich network representations.

\section{Conclusion}
\label{section6}
In this paper, we propose a novel
text-rich graph neural network, namely TeKo, which effectively integrates both network structure and textual semantics with guidance from external knowledge for text-rich network representations.
In specific, we first augment the original text-rich network structure into a heterogeneous semantic network by incorporating informative entities and interactions among documents and entities.
We then leverage two types of external knowledge, structured triples and unstructured entity descriptions, to learn jointly entity representations for deep understanding of text semantics.
We further design a reciprocal propagation mechanism for the  augmented heterogeneous semantic network, which realizes a well balanced combination of network structure and textual semantics, ultimately improving the quality of text-rich network representations.
Extensive experiments demonstrate the superior performance of the proposed new approach over the state-of-the-arts on four public text-rich networks as well as a large-scale real-world e-commerce searching dataset. 

\bibliographystyle{ieeetr} 
\bibliography{ref}

\begin{thebibliography}{10}

\bibitem{GNNBook2022}
L.~Wu, P.~Cui, J.~Pei, and L.~Zhao, {\em Graph Neural Networks: Foundations,
  Frontiers, and Applications}.
\newblock Springer Singapore, 2022.

\bibitem{DBLP:conf/icml/Ma0KW019}
J.~Ma, P.~Cui, K.~Kuang, X.~Wang, and W.~Zhu, ``Disentangled graph
  convolutional networks,'' in {\em Proceedings of ICML}, vol.~97,
  pp.~4212--4221, 2019.

\bibitem{DBLP:journals/corr/LiTBZ15}
Y.~Li, D.~Tarlow, M.~Brockschmidt, and R.~S. Zemel, ``Gated graph sequence
  neural networks,'' in {\em Proceedings of ICLR}, 2016.

\bibitem{DBLP:conf/ijcai/HeSJ0ZYZ20}
D.~He, Y.~Song, D.~Jin, Z.~Feng, B.~Zhang, Z.~Yu, and W.~Zhang,
  ``Community-centric graph convolutional network for unsupervised community
  detection,'' in {\em Proceedings of IJCAI}, pp.~3515--3521, 2020.

\bibitem{DBLP:conf/www/TanLZYZH20}
Q.~Tan, N.~Liu, X.~Zhao, H.~Yang, J.~Zhou, and X.~Hu, ``Learning to hash with
  graph neural networks for recommender systems,'' in {\em Proceedings of WWW},
  pp.~1988--1998, 2020.

\bibitem{DBLP:conf/kdd/ZhangSHSC19}
C.~Zhang, D.~Song, C.~Huang, A.~Swami, and N.~V. Chawla, ``Heterogeneous graph
  neural network,'' in {\em Proceedings of SIGKDD}, pp.~793--803, 2019.

\bibitem{DBLP:conf/icml/Abu-El-HaijaPKA19}
S.~Abu{-}El{-}Haija, B.~Perozzi, A.~Kapoor, N.~Alipourfard, K.~Lerman,
  H.~Harutyunyan, G.~V. Steeg, and A.~Galstyan, ``Mixhop: Higher-order graph
  convolutional architectures via sparsified neighborhood mixing,'' in {\em
  Proceedings of ICML}, vol.~97, pp.~21--29, 2019.

\bibitem{DBLP:conf/wsdm/JinSYLZC021}
D.~Jin, X.~Song, Z.~Yu, Z.~Liu, H.~Zhang, Z.~Cheng, and J.~Han, ``Bite-gcn: {A}
  new {GCN} architecture via bidirectional convolution of topology and features
  on text-rich networks,'' in {\em Proceedings of WSDM}, pp.~157--165, 2021.

\bibitem{DBLP:conf/icdm/YuJLHWT021}
Z.~Yu, D.~Jin, Z.~Liu, D.~He, X.~Wang, H.~Tong, and J.~Han, ``{AS-GCN:}
  adaptive semantic architecture of graph convolutional networks for text-rich
  networks,'' in {\em Proceedings of ICDM}, pp.~837--846, 2021.

\bibitem{DBLP:conf/aaai/SpeerCH17}
R.~Speer, J.~Chin, and C.~Havasi, ``Conceptnet 5.5: An open multilingual graph
  of general knowledge,'' in {\em Proceedings of AAAI}, pp.~4444--4451, 2017.

\bibitem{DBLP:conf/aaai/LinLSLZ15}
Y.~Lin, Z.~Liu, M.~Sun, Y.~Liu, and X.~Zhu, ``Learning entity and relation
  embeddings for knowledge graph completion,'' in {\em Proceedings of AAAI},
  pp.~2181--2187, 2015.

\bibitem{DBLP:conf/coling/SunSQGHHZ20}
T.~Sun, Y.~Shao, X.~Qiu, Q.~Guo, Y.~Hu, X.~Huang, and Z.~Zhang, ``Colake:
  Contextualized language and knowledge embedding,'' in {\em Proceedings of
  COLING}, pp.~3660--3670, 2020.

\bibitem{DBLP:conf/aaai/HuangXXDXLBXLY21}
C.~Huang, H.~Xu, Y.~Xu, P.~Dai, L.~Xia, M.~Lu, L.~Bo, H.~Xing, X.~Lai, and
  Y.~Ye, ``Knowledge-aware coupled graph neural network for social
  recommendation,'' in {\em Proceedings of AAAI}, pp.~4115--4122, 2021.

\bibitem{DBLP:conf/nips/ZhuYZHAK20}
J.~Zhu, Y.~Yan, L.~Zhao, M.~Heimann, L.~Akoglu, and D.~Koutra, ``Beyond
  homophily in graph neural networks: Current limitations and effective
  designs,'' in {\em Proceedings of NeurIPS}, 2020.

\bibitem{DBLP:conf/cikm/FerraginaS10}
P.~Ferragina and U.~Scaiella, ``{TAGME:} on-the-fly annotation of short text
  fragments (by wikipedia entities),'' in {\em Proceedings of CIKM},
  pp.~1625--1628, 2010.

\bibitem{DBLP:conf/iclr/KipfW17}
T.~N. Kipf and M.~Welling, ``Semi-supervised classification with graph
  convolutional networks,'' in {\em proceedings of ICLR}, 2017.

\bibitem{DBLP:conf/aaai/ZhuR0MLAK21}
J.~Zhu, R.~A. Rossi, A.~Rao, T.~Mai, N.~Lipka, N.~K. Ahmed, and D.~Koutra,
  ``Graph neural networks with heterophily,'' in {\em Proceedings of AAAI},
  pp.~11168--11176, 2021.

\bibitem{DBLP:conf/nips/BordesUGWY13}
A.~Bordes, N.~Usunier, A.~Garc{\'{\i}}a{-}Dur{\'{a}}n, J.~Weston, and
  O.~Yakhnenko, ``Translating embeddings for modeling multi-relational data,''
  in {\em Proceedings of NeurIPS}, pp.~2787--2795, 2013.

\bibitem{DBLP:journals/jmlr/BleiNJ03}
D.~M. Blei, A.~Y. Ng, and M.~I. Jordan, ``Latent dirichlet allocation,'' {\em
  Journal of Machine Learning Research}, vol.~3, pp.~993--1022, 2003.

\bibitem{DBLP:conf/ijcai/XuQCH17}
J.~Xu, X.~Qiu, K.~Chen, and X.~Huang, ``Knowledge graph representation with
  jointly structural and textual encoding,'' in {\em Proceedings of IJCAI},
  pp.~1318--1324, 2017.

\bibitem{DBLP:conf/emnlp/HuYSJL19}
L.~Hu, T.~Yang, C.~Shi, H.~Ji, and X.~Li, ``Heterogeneous graph attention
  networks for semi-supervised short text classification,'' in {\em Proceedings
  of EMNLP}, pp.~4820--4829, 2019.

\bibitem{DBLP:conf/iclr/VelickovicCCRLB18}
P.~Velickovic, G.~Cucurull, A.~Casanova, A.~Romero, P.~Li{\`{o}}, and
  Y.~Bengio, ``Graph attention networks,'' in {\em Proceedings of ICLR}, 2018.

\bibitem{DBLP:conf/kdd/TangZYLZS08}
J.~Tang, J.~Zhang, L.~Yao, J.~Li, L.~Zhang, and Z.~Su, ``Arnetminer: extraction
  and mining of academic social networks,'' in {\em Proceedings of SIGKDD},
  pp.~990--998, 2008.

\bibitem{DBLP:conf/icml/WuSZFYW19}
F.~Wu, A.~H.~S. Jr., T.~Zhang, C.~Fifty, T.~Yu, and K.~Q. Weinberger,
  ``Simplifying graph convolutional networks,'' in {\em Proceedings of ICML},
  vol.~97, pp.~6861--6871, 2019.

\bibitem{DBLP:conf/iclr/VelickovicFHLBH19}
P.~Velickovic, W.~Fedus, W.~L. Hamilton, P.~Li{\`{o}}, Y.~Bengio, and R.~D.
  Hjelm, ``Deep graph infomax,'' in {\em Proceedings of ICLR}, 2019.

\bibitem{DBLP:conf/www/PengHLZRXH20}
Z.~Peng, W.~Huang, M.~Luo, Q.~Zheng, Y.~Rong, T.~Xu, and J.~Huang, ``Graph
  representation learning via graphical mutual information maximization,'' in
  {\em Proceedings of WWW}, pp.~259--270, 2020.

\bibitem{DBLP:conf/nips/HamiltonYL17}
W.~L. Hamilton, Z.~Ying, and J.~Leskovec, ``Inductive representation learning
  on large graphs,'' in {\em Proceedings of NeurIPS}, pp.~1024--1034, 2017.

\bibitem{DBLP:conf/kdd/0017ZB0SP20}
X.~Wang, M.~Zhu, D.~Bo, P.~Cui, C.~Shi, and J.~Pei, ``{AM-GCN:} adaptive
  multi-channel graph convolutional networks,'' in {\em Proceedings of SIGKDD},
  pp.~1243--1253, 2020.

\bibitem{DBLP:conf/iclr/PeiWCLY20}
H.~Pei, B.~Wei, K.~C. Chang, Y.~Lei, and B.~Yang, ``Geom-gcn: Geometric graph
  convolutional networks,'' in {\em Proceedings of ICLR}, 2020.

\bibitem{DBLP:conf/emnlp/PenningtonSM14}
J.~Pennington, R.~Socher, and C.~D. Manning, ``Glove: Global vectors for word
  representation,'' in {\em Proceedings of EMNLP}, pp.~1532--1543, 2014.

\bibitem{DBLP:journals/tacl/WangGZZLLT21}
X.~Wang, T.~Gao, Z.~Zhu, Z.~Zhang, Z.~Liu, J.~Li, and J.~Tang, ``{KEPLER:} {A}
  unified model for knowledge embedding and pre-trained language
  representation,'' {\em Transactions of the Association for Computational
  Linguistics}, vol.~9, pp.~176--194, 2021.

\bibitem{29}
V.~D.~M. Laurens and G.~Hinton, ``Visualizing data using t-sne,'' {\em Journal
  of Machine Learning Research}, vol.~9, no.~2605, pp.~2579--2605, 2008.

\bibitem{mv-lstm}
S.~Wan, Y.~Lan, J.~Guo, J.~Xu, L.~Pang, and X.~Cheng, ``A deep architecture for
  semantic matching with multiple positional sentence representations.,'' in
  {\em Proceedings of AAAI}, pp.~2835--2841, 2016.

\bibitem{k-nrm}
C.~Xiong, Z.~Dai, J.~Callan, Z.~Liu, and R.~Power, ``End-to-end neural ad-hoc
  ranking with kernel pooling,'' in {\em Proceedings of SIGIR}, pp.~55--64,
  2017.

\bibitem{arc}
B.~Hu, Z.~Lu, H.~Li, and Q.~Chen, ``Convolutional neural network architectures
  for matching natural language sentences,'' in {\em Proceedings of NeurIPS},
  pp.~2042--2050, 2014.

\bibitem{matchpy}
L.~Pang, Y.~Lan, J.~Guo, J.~Xu, S.~Wan, and X.~Cheng, ``Text matching as image
  recognition.,'' in {\em Proceedings of AAAI}, vol.~16, pp.~2793--2799, 2016.

\bibitem{duet}
B.~Mitra, F.~Diaz, and N.~Craswell, ``Learning to match using local and
  distributed representations of text for web search,'' in {\em Proceedings of
  WWW}, pp.~1291--1299, 2017.

\bibitem{DBLP:conf/icdm/JiangSLSXXXYJ20}
Y.~Jiang, Y.~Shang, Z.~Liu, H.~Shen, Y.~Xiao, W.~Xiong, S.~Xu, W.~Yan, and
  D.~Jin, ``{BERT2DNN:} {BERT} distillation with massive unlabeled data for
  online e-commerce search,'' in {\em Proceedings of ICDM}, pp.~212--221, 2020.

\bibitem{DBLP:journals/corr/BrunaZSL13}
J.~Bruna, W.~Zaremba, A.~Szlam, and Y.~LeCun, ``Spectral networks and locally
  connected networks on graphs,'' in {\em Proceedings of ICLR}, 2014.

\bibitem{1}
M.~Defferrard, X.~Bresson, and P.~Vandergheynst, ``Convolutional neural
  networks on graphs with fast localized spectral filtering,'' in {\em
  Proceedings of NeurIPs}, pp.~3837--3845, 2016.

\bibitem{DBLP:conf/cikm/ShiSLZHLZW0019}
Y.~Shi, J.~Shen, Y.~Li, N.~Zhang, X.~He, Z.~Lou, Q.~Zhu, M.~Walker, M.~Kim, and
  J.~Han, ``Discovering hypernymy in text-rich heterogeneous information
  network by exploiting context granularity,'' in {\em Proceedings of CIKM},
  pp.~599--608, 2019.

\bibitem{DBLP:conf/www/ShangZLL020}
J.~Shang, X.~Zhang, L.~Liu, S.~Li, and J.~Han, ``Nettaxo: Automated topic
  taxonomy construction from text-rich network,'' in {\em Proceedings of WWW},
  pp.~1908--1919, 2020.

\bibitem{DBLP:conf/www/ZhangZDS021}
X.~Zhang, C.~Zhang, X.~L. Dong, J.~Shang, and J.~Han, ``Minimally-supervised
  structure-rich text categorization via learning on text-rich networks,'' in
  {\em Proceedings of WWW}, pp.~3258--3268, 2021.

\bibitem{DBLP:conf/www/WangZXLG19}
H.~Wang, M.~Zhao, X.~Xie, W.~Li, and M.~Guo, ``Knowledge graph convolutional
  networks for recommender systems,'' in {\em Proceedings of WWW},
  pp.~3307--3313, 2019.

\bibitem{DBLP:conf/kdd/Wang00LC19}
X.~Wang, X.~He, Y.~Cao, M.~Liu, and T.~Chua, ``{KGAT:} knowledge graph
  attention network for recommendation,'' in {\em Proceedings of SIGKDD},
  pp.~950--958, 2019.

\bibitem{DBLP:conf/aaai/ZhangZZ0XH20}
Z.~Zhang, F.~Zhuang, H.~Zhu, Z.~Shi, H.~Xiong, and Q.~He, ``Relational graph
  neural network with hierarchical attention for knowledge graph completion,''
  in {\em Proceedings of AAAI}, pp.~9612--9619, 2020.

\bibitem{DBLP:conf/cikm/LiLZHWYW20}
J.~Li, S.~Li, W.~X. Zhao, G.~He, Z.~Wei, N.~J. Yuan, and J.~Wen,
  ``Knowledge-enhanced personalized review generation with capsule graph neural
  network,'' in {\em Proceedings of CIKM}, pp.~735--744, 2020.

\bibitem{DBLP:conf/acl/HuYZZTSD020}
L.~Hu, T.~Yang, L.~Zhang, W.~Zhong, D.~Tang, C.~Shi, N.~Duan, and M.~Zhou,
  ``Compare to the knowledge: Graph neural fake news detection with external
  knowledge,'' in {\em Proceedings of ACL}, pp.~754--763, 2021.

\bibitem{DBLP:conf/aaai/YaoM019}
L.~Yao, C.~Mao, and Y.~Luo, ``Graph convolutional networks for text
  classification,'' in {\em Proceedings of AAAI}, pp.~7370--7377, 2019.

\bibitem{DBLP:conf/aaai/LiuYZWL20}
X.~Liu, X.~You, X.~Zhang, J.~Wu, and P.~Lv, ``Tensor graph convolutional
  networks for text classification,'' in {\em Proceedings of AAAI},
  pp.~8409--8416, 2020.

\bibitem{DBLP:conf/acl/ZhangYCWWW20}
Y.~Zhang, X.~Yu, Z.~Cui, S.~Wu, Z.~Wen, and L.~Wang, ``Every document owns its
  structure: Inductive text classification via graph neural networks,'' in {\em
  Proceedings of ACL}, pp.~334--339, 2020.

\bibitem{DBLP:conf/ijcai/LiPPLW21}
C.~Li, X.~Peng, H.~Peng, J.~Li, and L.~Wang in {\em Proceedings of IJCAI},
  pp.~2680--2686, 2021.

\end{thebibliography}

\end{document}